\documentclass[a4paper,fleqn,usenatbib]{mnras}
\usepackage[T1]{fontenc}
\usepackage{ae,aecompl}

\usepackage{graphicx}	
\usepackage{amsmath}	
\usepackage{amssymb}	
\usepackage{pdflscape}

\title[Gaia Stellar Kinematics]{Gaia Stellar Kinematics in the Head of the Orion A Cloud: Runaway Stellar Groups and Gravitational Infall}

\author[K. V. Getman et al.]{
K. V. Getman,$^{1}$\thanks{E-mail: kug1@psu.edu (KVG)}
E. D. Feigelson,$^{1}$
M. A. Kuhn,$^{2}$
G. P. Garmire$^{3}$
\\
$^{1}$Department of Astronomy \& Astrophysics, 525 Davey Laboratory, Pennsylvania State University, University Park PA 16802\\
$^{2}$Department of Astronomy, California Institute of Technology, Pasadena, CA 91125, USA\\
$^{3}$Huntingdon Institute for X-ray Astronomy, LLC, 10677 Franks Road, Huntingdon, PA 16652, USA
}

\date{Accepted for publication in MNRAS; 05/23/19}

\pubyear{2018}

\begin{document}
\label{firstpage}
\pagerange{\pageref{firstpage}--\pageref{lastpage}}
\maketitle

\begin{abstract}
	This work extends previous kinematic studies of young stars in the Head of the Orion A cloud (OMC-1/2/3/4/5). It is based on large samples of infrared, optical, and X-ray selected pre-main sequence  stars with reliable radial velocities and {\it Gaia}-derived parallaxes and proper motions. Stellar kinematic groups are identified assuming they mimic the motion of their parental gas. Several groups are found to have peculiar kinematics: the NGC~1977 cluster and two stellar groups in the Extended Orion Nebula (EON) cavity are caught in the act of departing their birthplaces.  The abnormal motion of NGC~1977 may have been caused by a global hierarchical cloud collapse, feedback by massive Ori OB1ab stars, supersonic turbulence, cloud-cloud collision, and/or slingshot effect; the former two models are favored by us. EON groups might have inherited anomalous motions of their parental cloudlets due to small-scale `rocket effects' from nearby OB stars.  We also identify sparse stellar groups to the east and west of Orion A that are drifting from the central region, possibly a slowly expanding halo of the Orion Nebula Cluster.  We confirm previously reported findings of varying line-of-sight distances to different parts of the cloud's Head with associated differences in gas velocity. Three-dimensional movies of star kinematics show contraction of the groups of stars in OMC-1 and global contraction of OMC-123 stars.  Overall, the Head of Orion A region exhibits complex motions consistent with theoretical models involving hierarchical gravitational collapse in (possibly turbulent) clouds with OB stellar feedback.
\end{abstract}




\section{Introduction} \label{sec_introduction}

Shells, bubbles, and filamentary molecular cloud structures are ubiquitous in the Galaxy \citep{Churchwell2006,Andre2014} and are often sites of star formation. Hierarchical fragmentation in molecular cloud filaments is often observed on scales ranging from several parsecs to $\la 0.1$~pc. Different physical mechanisms appear to trigger the cloud fragmentation including gravitational collapse; thermal, turbulent, and magnetic pressures; angular momentum; and dynamical feedback from young stellar outflows, winds, and radiation pressure \citep{Takahashi2013,Contreras2016,Teixeira2016}. It is also possible that clouds are largely composed of velocity-coherent sub-filaments \citep{Hacar2013,Hacar2018}. Turbulent energy cascades or global hierarchical gravitational collapse enhancing anisotropies are proposed to play major roles in the formation of both the sub-filaments and integrated filaments \citep[e.g.,][]{Smith2016,Vazquez-Semadeni2019}. Dense, gravitationally bound pre-stellar cores then form by cloud fragmentation along the densest filaments; core growth through filamentary accretion is also reported \citep{Andre2014}. Small star clusters may emerge in these cores through star formation mediated by  turbulent core accretion \citep{McKee2003}, competitive accretion \citep{Bonnell2001,Wang2010}, stellar mergers \citep{Bonnell2005}, and/or global hierarchical gravitational collapse \citep{Vazquez-Semadeni2017,Vazquez-Semadeni2019}. Molecular gas can then be expelled by feedback effects of young stars including OB ionizing radiation and winds, supernovae, protostellar accretion heating, protostellar jets and outflows \citep{Dale2015}.

The Orion A and B giant molecular filaments are the two most prominent sites of recent active star formation \citep{Megeath2012, Meingast2016} within the nearby and richest complex of low- and high-mass young stars, the Orion Complex \citep{Bally2008}. Our current study focuses on the distance and kinematics properties of young stars located in the northern part of Orion A. Following \citet{Grossschedl2018a}, here we use the ``Head'' and ``Tail'' designations for the northern and southern parts of this cometary shaped cloud, respectively \citep{Bally1987}. Harboring the massive Orion Nebula Cluster (ONC), the Head (formerly called the Integral Shaped Filament) is composed of several molecular components: OMC-2/3, OMC-1, OMC-4, and OMC-5 \citep{Bally1987,Johnstone1999,Johnstone2006,Wu2018}.

Several research groups report that the stellar and gas radial velocities across Orion A are in a remarkable agreement with each other \citep{Furesz2008,Tobin2009,Hacar2016,Kounkel2016,DaRio2017}. The stellar velocities closely follow the large-scale north-south velocity gradient of the gas across entire Orion A cloud, with velocities ranging from about $V_{rad,LSR} \ga 12$~km~s$^{-1}$ at the northern tip down to $V_{rad,LSR} \la 2$~km~s$^{-1}$ at the southern tip of the cloud. Based on this finding \citet{Furesz2008} propose that the bulk of young stars in the Head of Orion A still mimic the motion of their parental molecular material. \citet{Hacar2016} further report the presence of strings of stars exhibiting low-velocity dispersions, characteristic of their parental gas. Considering the recent observational findings of gas inflall in the Head of the cloud \citep[e.g.,][]{Hacar2017,Wu2018}, it would be reasonable to expect for bulk of young stars in the Head to be in a dynamical state of contraction. 

However, recent studies of star kinematics in Orion A find no evidence for star contraction. \citet{DaRio2017} provide signs of star expansion in the ONC based on the correlation between source extinction and radial velocities derived via near-IR Apache Point Observatory Galactic Evolution Experiment (APOGEE) spectroscopy.  The {\it Gaia}-APOGEE study of the Orion Complex including Orion A by \citet{Kounkel2018} reports random stellar motions across ONC with ``slight preference for expansion near the outer edges''. Our recent {\it Gaia} study (with no consideration of the radial velocity component) of numerous star forming regions by \cite{Kuhn2018} finds the majority of the studied young stellar clusters and associations in a state of dynamical expansion, with some evidence for mild expansion and gravitational boundedness of the ONC. The HST/Keck study of stellar proper motions in the central ONC by \citet{Kim2018} finds no signs of ONC expansion. Kinematic analyses of stars in Orion A are further affected by uncertain distance measurements towards the Head of the cloud. For instance, \citet{Kuhn2018} and \citet{Stutz2018b} report varying {\it Gaia} distances while \citet{Kounkel2018} and \citet{Grossschedl2018a} derive constant {\it Gaia} distances towards different parts of the Head.

The main purpose of this study is to use rich samples of young stars in Orion A obtained with X-ray, infrared, and optical surveys, combined with reliable radial velocities and {\it Gaia}-derived parallaxes and proper motions, to search for signs of stellar motion $-$ contraction, expansion, and anomalous flows $-$ across the cloud's Head. Our sample size of young stars exceeds those of \citet{Kuhn2018,Stutz2018b,Grossschedl2018a}, and our selection of kinematic stellar structures across the Head is conceptually different from that of \citet{Kounkel2018}. Our analyses find star-gas contraction across the cloud's Head as well as lead to a serendipitous discovery of one stellar cluster and several stellar groups with peculiar kinematics.

The paper is organized as follows. Stellar samples and kinematic methodology are reviewed in \S \ref{sec_methods}. Our selection of stellar kinematic structures is given in \S\ref{sec_stellargroups}. Detailed analyses of {\it Gaia} distances are given in \S \ref{sec_distance}. The discovery of stellar cluster/groups with peculiar kinematics is presented in \S\S \ref{sec_ngc1977} and \ref{sec_eastweststars}. Findings on star-gas contraction are provided in \S\S \ref{sec_3dmovie} and \ref{sec_relationship_d_vs_vr}. Section \ref{sec_discussion} discusses the implications of our findings for the star formation in the Orion A filament. The Appendices provide additional material, including catalog tables (\S \ref{appendix_3tables}), velocity transformation procedures (\S \ref{appendix_v_transformation}), raw spatial maps of distance and velocities (\S\ref{appendix_raw_maps}), and raw and adaptively smoothed maps of stellar ages across the cloud's Head (\S \ref{appendix_b}).

\section{Samples And Methods}\label{sec_methods}

\subsection{{\it Gaia} Star Sample Selection}\label{sec_starsamples}

To produce a rich sample of {\it Gaia} young stars in the Orion A region we combine data from several different young stellar catalogues. We utilize X-ray/IR data from the Star Formation in Nearby Clouds project \citep[SFiNCs;][]{Getman2017} and the {\it Chandra} Orion Ultradeep Project \citep[COUP;][]{Getman2005}; the latter is part of the Massive Young Star-Forming Complex Study in Infrared and X-ray \citep[MYStIX;][]{Feigelson2013,Broos2013}. The COUP-MYStIX observations, centered on the Trapezium core of the ONC cluster, cover about $17 \times 17$~arcmin$^{2}$ area on the sky \citep[][their Figure 2]{Getman2005}. The SFiNCs observations encompass the NGC~1977, OMC-2/3, and OMC-4 regions \citep[][their Figure Set 1]{Getman2017}, covering three $17 \times 17$~arcmin$^{2}$ and three roughly $20 \times 30$~arcmin$^{2}$ areas with {\it Chandra} and {\it Spitzer}, respectively. We add optical/IR young stars from \citet{Furesz2008} and \citet{Tobin2009} that were selected on the basis of their 2MASS, {\it Spitzer}-IRAC, radial velocity, and H$\alpha$ emission properties. These catalogs cover the $83.4 < \alpha < 84.3\degr$ and $-6.6 < \delta < -4.4\degr$ stripe on the sky (roughly $55 \times 130$~armin$^{2}$) encompassing the NGC~1977, OMC-1/2/3/4/5, and L1641-N regions \citep[][their Figure 1]{Tobin2009}. We also incorporate the catalog of young stars from \citet{DaRio2012}, selected on the basis of their optical/IR photometry and spectroscopy; these lie within the area of roughly $33 \times 33$~armin$^{2}$ enclosing OMC-1 and part of OMC-4 \citep[][their Figure~15]{DaRio2012}. The catalog of young stars from \citet{DaRio2016} is added; it is composed of IR-bright stars whose youthfulness was supported by various IR/optical/X-ray indicators from the literature as well as by a followup near-IR APOGEE spectroscopy. This catalog covers the entire Orion A cloud, approximately within the sky band of $\Delta \alpha = 1.1\degr$ and $-9.7 < \delta < -4.1\degr$ \citep[][their Figure 19]{DaRio2016}. We also include the {\it Spitzer}-selected disky young stellar objects (YSOs) from \citet{Megeath2012}; its coverage of Orion A is roughly similar to that of the \citet{DaRio2016} catalog \citep[][their Figure~9]{Megeath2012}. Finally, the large VISTA  catalog of near-IR $JHK_s$ \citep{Meingast2016} and mid-IR {\it Spitzer}-IRAC \citep{Megeath2012} point sources covering the entire Orion A are accessed. 

For each of the above catalogs, we performed cross-correlations between {\it Gaia} DR2 \citep{Gaia2018} and catalog source positions within a constant search radius of $1\arcsec$. Duplicate {\it Gaia} sources are identified and removed. The {\it Gaia} sample is then restricted to stars with statistical uncertainty on parallax $\sigma_{\bar{\omega}}<0.1$~mas; this criterion permits astrometric excess noise $\epsilon_{i}<0.5$~mas \citep{Lindegren2012} and reliable kinematics ($\sigma_{\mu_{\alpha\star}} < 0.18$~mas~yr$^{-1}$, $\sigma_{\mu_{\delta}} < 0.18$~mas~yr$^{-1}$) for 98\% of the stars. Furthermore, only stars with parallax measurements in the range $2<\bar{\omega}<3$~mas ($2.2<\bar{\omega}<2.8$~mas) are retained for the analysis of distance across the entire Orion A (distance/kinematics across the Head). The resulting {\it Gaia} catalog of 1487 young stars with $2<\bar{\omega}<3$~mas is listed in Tables~\ref{tbl_gaia_props} and \ref{tbl_other_props}. These stars span a wide range of $G$-band magnitude, from 6.5 to 17.2~mag, with a skewed distribution peaking at $15.5-16$~mag. The statistical error on $G$ does not exceed 0.08~mag.

\begin{figure*}
	\includegraphics[angle=0.,width=170mm]{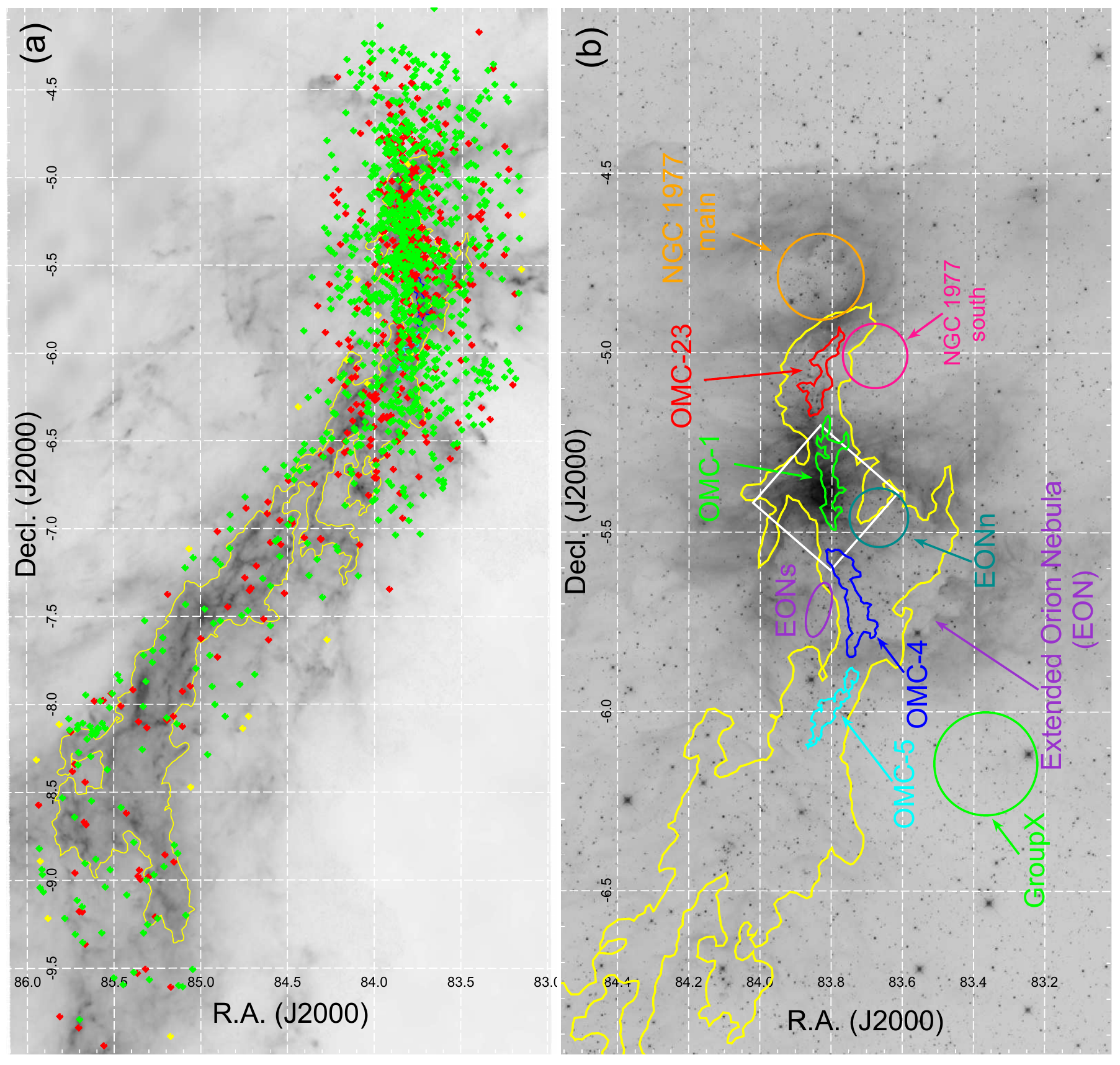}
	\caption{(a) Spatial distribution of 1487 {\it Gaia}-selected young stars in Orion A with parallax error $\sigma_{\bar{\omega}}<0.1$~mas. Symbols indicate 978 diskless (green), 470 disky (red), and 39 unclassified (yellow) stars superimposed on the {\it Herschel}-{\it Planck} dust optical-depth map of Orion A from \citet{Lombardi2014}. The yellow contour indicates the dust extinction at $A_V = 6$~mag. (b) Closeup of the Head of Orion A showing colored circles and contours for various stellar clustered components and molecular structures discussed in this paper.  They are superimposed on the WISE-3.4~$\mu$m image. The MYStIX-COUP field of view from \citet{Getman2005} (white square) and the $A_V = 6$~mag contour from Lombardi et al. (yellow contour) are provided. All further maps are restricted to the Orion Head region. }
	\label{fig_lombardi_map}
\end{figure*}

Table~\ref{tbl_gaia_props} focuses on {\it Gaia} properties, such as star positions, parallax ($\bar{\omega}$), proper motions in right ascension ($\mu_{\alpha\star}$) and declination ($\mu_{\delta}$), and $G$-band magnitude. The last column of the table gives a flag, FEC (flux-excess-cut), indicating photometric BP/RP flux excess: FEC=1 and FEC=0 select 1221 and 266 stars with small and large BP/RP flux excesses, respectively. According to \citet{Evans2018}, high excess in the sum of the $BP$-band and $RP$-band fluxes relative to the $G$-band flux may indicate that star's photometry is susceptible to the effects of source crowding and/or nebulosity. These factors in turn may be linked to potential distortions of source's PSF shape possibly leading to a poor astrometric fit (Timo Prusti; private communication). Our distance and kinematic analyses are carried out using two samples of {\it Gaia} stars: all stars irrespective of the BP/RP flux excess, i.e., with FEC=0 and FEC=1 (\texttt{full}); and only stars with FEC=1 (\texttt{restricted}). Below, major distance/kinematic results for the \texttt{full} and \texttt{restricted} samples appear qualitatively similar. Figures related to the \texttt{full} and \texttt{restricted} permutations are given in the main paper and Supplementary materials, respectively.   

Table~\ref{tbl_other_props} presents several stellar properties taken from previous literature. These include optical/IR/X-ray YSO-selection flag, near-IR $J$ and $H$-band magnitudes, visual source extinction ($A_V$), two age estimates based on traditional HRD methods and X-ray/near-IR photometry (the latter is termed $Age_{JX}$), apparent mid-IR SED (spectral energy distribution) slope, and a flag indicating membership in stellar structures considered in this paper. Details on these quantities are given in Appendix \S \ref{appendix_3tables}.

Following the study of disk lifetimes across numerous MYStIX/SFiNCs clusters by \citet{Richert2018}, we opt for a simple scheme, based on the apparent IRAC SED slope $\alpha_{IRAC}$ (Table~\ref{tbl_other_props}), to separate the {\it Gaia} stars into disky and diskless. Stars with $\alpha_{IRAC}$ above (below) the threshold value of -1.9 are classified as disky (diskless). Such classification yields 470 disky, 978 diskless, and 39 unclassified {\it Gaia} stars. Figure \ref{fig_lombardi_map}a shows the spatial distribution of all {\it Gaia} young stars stratified by the presence of disk superimposed on the {\it Herschel}-{\it Planck} dust optical-depth map of Orion A from \citet{Lombardi2014}. Clearly, both disky and diskless {\it Gaia} stars are prevalent across the entire Orion A cloud. As the richness of our {\it Gaia} sample decreases drastically from the Head to the Tail of the cloud, we restrict our kinematic analysis to the Head with $\delta > -6.5\degr$. 

Figure \ref{fig_lombardi_map}b shows a closeup view of the cloud's Head in the WISE 3.4~$\mu$m band, emphasizing the main cloud and stellar components utilized in our kinematic analyses below. Four main cloud components $-$ OMC-2/3, OMC-1, OMC-4, and OMC-5 $-$ are selected from high-resolution CARMA-NRO C$^{18}$O intensity and $^{13}$CO and C$^{18}$O first-moment maps from \citet{Kong2018}.  We identify eleven stellar components: four associated with the main cloud components, five related to relatively compact stellar groups/clusters with peculiar kinematics (denoted here as NGC 1977 main cluster, NGC 1977 south group, EONn and EONs groups, and Group~X), and two sparse stellar groups (not shown in the figure) lying to the east ($\alpha > 84\degr$) and west ($\alpha <83.6\degr$) of OMC-1/2/3/4. The procedure for selecting all these components is further detailed in \S \ref{sec_stellargroups}. Clearly, these stellar components are different from the kinematic stellar groups of \citet{Kounkel2018}.

Compared to recent {\it Gaia}-based stellar kinematic studies of the Orion A region (or its individual sub-regions), our {\it Gaia} sample size of 1487 young stars considerably exceeds the disky star samples of \citet[][682 stars]{Grossschedl2018a} and \citet[][500 stars]{Stutz2018b}, and the diskless/disky ONC sample of \citet[][378 stars]{Kuhn2018}.  This large sample is essential to our investigation of stellar motions in the Orion Head. Our sample size is roughly similar to that of \citet{Kounkel2018}\footnote{The selection of young stellar candidates in Table~2 of Kounkel et al. is solely based on constraining star's {\it Gaia} properties (such as positions, parallaxes, proper motions, and magnitudes). We count in total 1701 {\it Gaia} stars listed in Table~2 of Kounkel et al. that are identified as members of 57 stellar groups in/around the Orion A cloud; of those 1362 lie  inside the field covered by our {\it Gaia}-selected sample of 1487 young stars (Table~\ref{tbl_gaia_props}).}, but the young star/group selection methods and distance/kinematic procedures are conceptually different (\S~\ref{sec_stellargroups}).

\subsection{Stellar Radial Velocities}\label{sec_radial_velocities}

We utilize radial velocity measurements of point sources in Orion A using two publicly available datasets: data obtained with the near-IR APOGEE spectrograph on the 2.5~m Sloan Digital Sky Survey telescope \citep{Kounkel2018} complemented by optical spectroscopy data with Hectochelle on the MMT and MIKE Fibers on the Magellan Clay telescope \citep{Tobin2009}. As with the {\it Gaia}-selection above, here we opt to reduce the contamination from field stars in the Kounkel et al. catalog by selecting only sources that have $1\arcsec$-separation counterparts in the aforementioned optical/IR/X-ray YSO catalogs (\S\ref{sec_starsamples}). This source selection yields 2752 stars: 2326 bright (down to $H \sim 12.5$~mag) young stars across Orion A with APOGEE-derived radial velocities and 426 additional fainter (down to $H \sim 14$~mag) young stars in the Head of Orion A with velocity measurements provided solely by Tobin et al. For all these stars, Table~\ref{tbl_vrad} gives heliocentric average stellar velocity measurements from Kounkel et al. and Tobin et al. along with near-IR $J$ and $H$-band magnitudes (see Appendix \S\ref{appendix_3tables} for details).

For the analyses of stellar kinematics across the Head of Orion A provided below, we employ two subsamples of stars. The first subsample, denoted hereafter as ``entire K18T09'', is composed of 1715 stars that have radial velocities within the range $10<V_{rad}<40$~km~s$^{-1}$, lie within the $-6.5 < \delta < -4.0\degr$ spatial stripe, and are not known to be spectroscopic binaries (Mult$=1$ in Table~\ref{tbl_vrad}). The second subsample, denoted as ``{\it Gaia} K18T09'', represents a fragment of the ``entire K18T09'' subsample further limited to 904 stars that have counterparts in our {\it Gaia}-selected young star catalog (Table~\ref{tbl_gaia_props}). Below, kinematic results for both subsamples appear similar; figures related to the ``entire K18T09''  and ``{\it Gaia} K18T09'' permutations are presented in the main paper and Supplementary materials, respectively.   

\citet{Kounkel2018} acknowledge that there is a systematic difference between their APOGEE-derived and previously reported stellar radial velocities. Our comparison of the radial velocity measurements for stars that are common between Kounkel et al. and \citet{Tobin2009} gives the following linear regression fit that treats variables symmetrically: $V_{rad,K18} = V_{rad,T09} \times 0.88 (\pm 0.5) + 4.10(\pm 1.33)$. Tobin et al. radial velocities are systematically lower by about $1$~km~s$^{-1}$ than those of Kounkel et al. For our study of star kinematics in the Head of Orion A, all velocity measurements from Tobin et al. have been corrected for this bias; Table~\ref{tbl_vrad} reports original uncorrected values.

\subsection{Stellar Kinematics}\label{sec_stellarkinematics}

All stellar kinematic analyses below are performed with respect to the rest frame of the star center in the Head of Orion A using a Cartesian $x,y,z$ coordinate system where the $x$ and $y$ axes are orthographic projections of the $-\alpha$ (i.e., direction opposite to the right ascension axis) and $+\delta$ (i.e, along the declination axis) celestial lines, and the $z$-axis is directed along the line-of-sight. Corresponding stellar velocities are denoted as $V_X$, $V_Y$, and $V_Z$.

The field center ($\alpha_0 = 83.820860\degr$, $\delta_0 = -5.4010458\degr$) is chosen as the median position for the {\it Gaia} stars in the Head of Orion A within the the $-6.5 < \delta < -4.0\degr$ and $83.6 < \alpha < 84\degr$ rectangle. It is located around $50\arcsec$ north of OMC-1S and about $50\arcsec$ south-east of $\Theta^1$C~Ori. Details regarding the transformation of the stellar proper motions and radial velocities to $V_X$, $V_Y$, and $V_Z$ are given in Appendix \ref{appendix_v_transformation}.

Individual stellar distances are obtained by inverting the {\it Gaia} parallaxes. Distances based on the probabilistic analysis of \citet{BailerJones2018} are offset from the above distances by $\sim 4$~pc across the entire Orion A cloud. Other recent studies on {\it Gaia} distances in Orion A similarly use inverses of parallaxes \citep{Kounkel2018, Grossschedl2018a, Kuhn2018, Stutz2018b}. Analysis of the stellar distances is  presented in \S~\ref{sec_distance}.

\subsection{Statistical Methods}\label{sec_statisticalmethods}

All statistical procedures in this paper were performed using the {\it R} statistical software environment \citep{Rcoreteam2018}, including several  \texttt{CRAN} packages. Throughout the paper local quadratic regression fits to bivariate datasets are generated using the {\it LOCFIT.robust} function from the \texttt{LOCFIT} package \citep{Loader1999,Loader2018}. Uncertainties on median values are estimated using non-parametric bootstrap re-sampling implemented by the functions {\it boot} and {\it boot.ci} from the \texttt{BOOT} package \citep{Davidson1997,Canty2017}.  Adaptive Gaussian kernel smoothing maps for point processes with a single mark variable are constructed using the algorithm {\it adaptive.density} from the package \texttt{SPATSTAT} \citep{Baddeley2015}. The non-parametric Cram\'er and k-sample Anderson-Darling tests used here to test the equality of multivariate and univariate data are performed using the {\it cramer.test} and {\it ad.test} functions from the \texttt{CRAMER} \citep{Carsten2014} and \texttt{KSAMPLES} \citep{Scholz2018} packages, respectively. The non-parametric Kendall's $\tau$ test employed here to establish associations between two variables is conducted using the {\it corr.test} function from the \texttt{PSYCH} package \citep{Revelle2018}. Symmetrical linear regressions are performed using the \texttt{lmodel2} package \citep{Legendre2018}. 2-D and 3-D data visualizations are presented here utilizing various functions from the \texttt{GGPLOT2} \citep{Wickham2018} and \texttt{PLOT3D} \citep{Soetaert2017} packages.

\section{Results}\label{sec_results}
\subsection{Stellar Kinematic Groups}\label{sec_stellargroups}

\begin{figure*}
	\includegraphics[angle=0.,width=150mm]{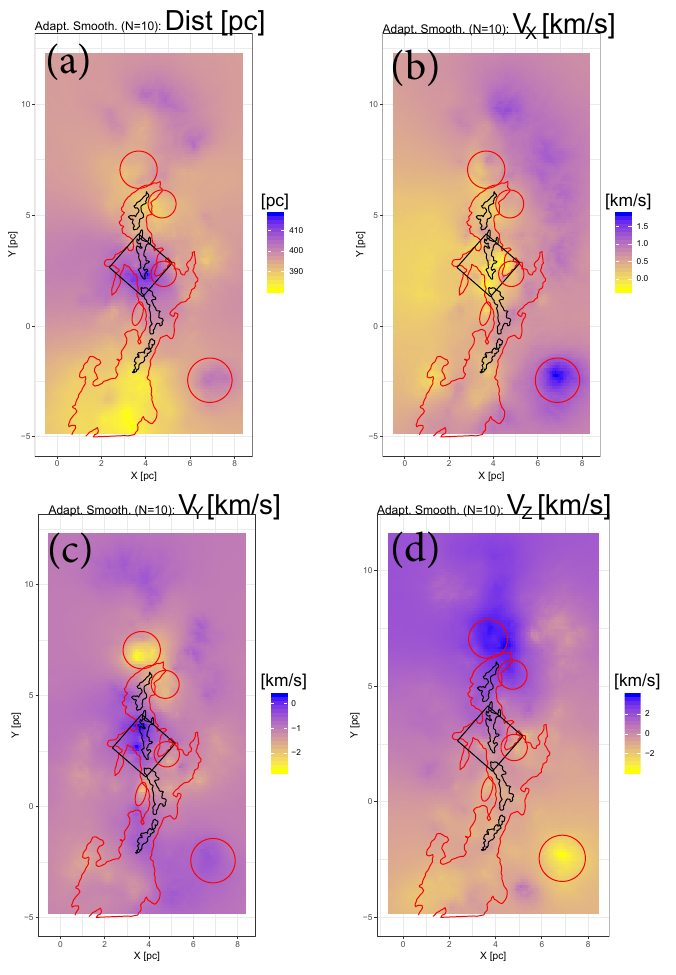}
	\caption{Spatial maps of adaptively smoothed stellar distributions of stellar distances $D$  and velocity components, $V_X$, $V_Y$, and $V_Z$ for the {\it Gaia}  \texttt{full} sample. Complementary figure panels providing raw maps of these quantities, i.e., showing each individual star, can be found in Appendix \ref{appendix_raw_maps}. The $V_Z$ measurements are given for the ``K18T09'' sample. Outliers with extreme values of $D$, $V_X$, $V_Y$, and $V_Z$ are excluded from these maps to shorten dynamic ranges and allow meaningful color scales. For the OMC-12345 gas components, the C$^{18}$O$(1-0)$ emission contours at 7~K~km~s$^{-1}$ from \citet{Kong2018} are in black; the NGC 1977, EON, and GroupX stellar clusters/groups are marked by the red circles/ellipses. As reference contours, the MYStIX-COUP field of view (black square) and the $A_V = 6$~mag contour (red) are also provided. On all panels, the three tiny red points mark the locations of $\Theta^{1}$Ori~C, and the approximate centers of the BN-KL and OMC-1S sub-regions. Figure panels showing the $D$, $V_X$, $V_Y$ maps for the {\it Gaia} \texttt{restricted} sample are provided in the Supplementary Materials. The Supplementary figure also includes a $V_Z$ panel for the ``{\it Gaia}-K18T09'' star sample.}
	\label{fig_smoothed_maps}
\end{figure*}

Adaptively smoothed maps of star positions and position-velocity (PV) diagrams, provide a clear visualization of the stellar kinematics and distance distributions across the Head of the Orion cloud.  Figure~\ref{fig_smoothed_maps} presents adaptively smoothed maps of the {\it Gaia} star distances and the three velocity components $V_X$, $V_Y$, and $V_Z$. Recall that the directions of $V_X$, $V_Y$, and $V_Z$ are $-\alpha$, $+\delta$, and line-of-sight, respectively. The smoothing is performed on the $x$-$y$ plane using an adaptive Gaussian kernel that encompasses ten nearest neighbors. A figure showing raw (individual star) distributions is further given in Appendix \ref{appendix_raw_maps}.  Several interesting features are evident:
\begin{enumerate}

\item In Figure~\ref{fig_smoothed_maps}a, the {\it Gaia} distances from the Sun are larger for the stars that lie projected against OMC-1 and OMC-4 (blue in the picture) compared to the stars around OMC-2/3 and OMC-5 (yellow in the picture).

\item In Figure~\ref{fig_smoothed_maps}b, stars to the west of the cloud have higher $V_X$ velocities (blue) than stars to the east of the cloud (yellow). 

\item In Figures~\ref{fig_smoothed_maps}c and d, the $V_Y$ and $V_Z$ motions of the stars at the northern tip of OMC-2/3 (denoted NGC~1977 main cluster and NGC~1977 south group [in Figure \ref{fig_lombardi_map}b]), are distinct from stars elsewhere in the region. 

\item The distance and kinematics of a stellar group located at the south-west corner of the field (denoted Group~X) are distinct from those of most of the stars in the region. 

\item In Figures~\ref{fig_smoothed_maps}c and d, the $V_Y$ and $V_Z$ motions of two stellar groups lying immediately to the west of OMC-1 and east of OMC-4, may be different from those of the bulk of nearby stars; these groups are further denoted as EON-north and EON-south groups. 

\item The well-established north-south $V_Z$ velocity gradient (discussed below) is apparent in Figure~\ref{fig_smoothed_maps}d. 
\end{enumerate}

\begin{figure*}
	\includegraphics[angle=0.,width=130mm]{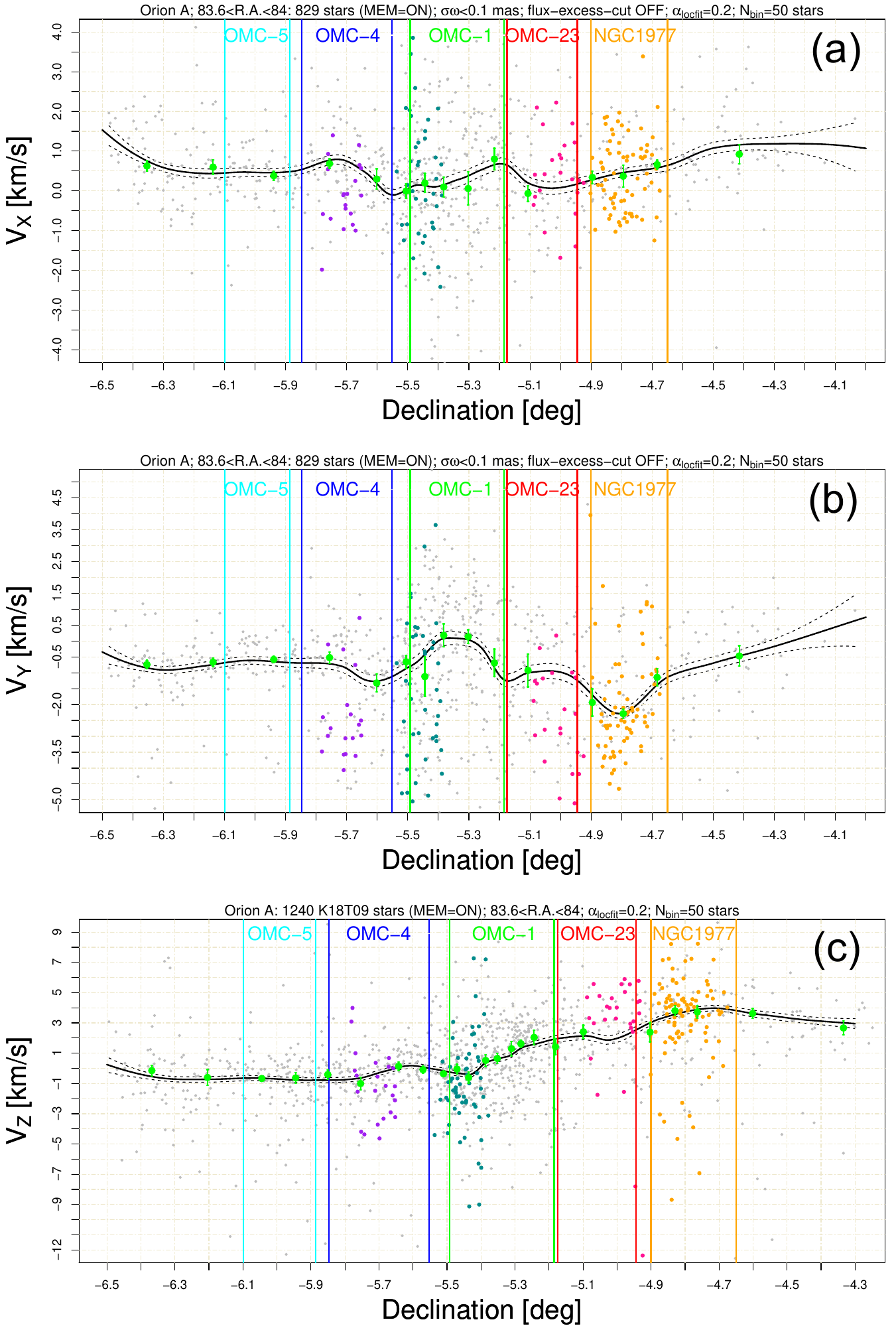}
	\caption{Transverse ($V_X$, $V_Y$) and radial ($V_Z$) velocities as functions of declination. On all panels, members of kinematically distinct stellar groups are color-coded: NGC~1977m (i.e., the main NGC~1977 stellar cluster; orange), NGC~1977s (i.e., a stellar group to the south of NGC~1977m; pink), EONn (dark green), and EONs (purple). The black solid and dashed curves indicate local quadratic regression fits and their $68$\% confidence intervals (CIs); the green points with error bars mark medians with their 68\% bootstrap uncertainties. The members of NGC~1977s, EONn, and EONs are omitted from the calculation of these local regression fits and median values. The orange, red, green, blue, and cyan vertical lines indicate the spatial boundaries of NGC~1977, OMC-23, OMC-1, OMC-4, and OMC-5, respectively. The current $V_X$ and $V_Y$ plots are presented for the {\it Gaia} \texttt{full} sample. Similar plots for the {\it Gaia} \texttt{restricted} sample are provided in the Supplementary Materials. The Supplementary figure also includes a $V_Z$ panel for the ``{\it Gaia}-K18T09'' star sample.}
	\label{fig_v_vs_declination}
\end{figure*}

Figure~\ref{fig_v_vs_declination} shows PV diagrams for the {\it Gaia}-selected (Table \ref{tbl_gaia_props}) and K18T09 (Table \ref{tbl_vrad}) young stars around the cloud's Head. The fitted curves depict trends of velocities as functions of declination; the cloud is conveniently elongated along this line (Figure~\ref{fig_lombardi_map}). The trends were obtained using two statistical procedures giving consistent results: local regression fits with \texttt{LOCFIT} (black) and running medians with bootstrap errors (green).  In accord with adaptive smoothed maps, the individual $V_Y$ and $V_Z$ PV positions of the NGC~1977m, NGC~1977s, EONn, and EONs stars (orange, pink, dark green, and purple) seem differ from the characteristic motions of stars in OMC-1/2/3/4/5.

In order to quantify characteristic motions of stars associated with OMC-1/2/3/4/5, with a desirably small contribution from the motions of the aforementioned nearby kinematic outlier stellar groups, the  members of the NGC~1977s (located next to OMC-23) and the EONn and EONs groups (located near OMC-1/4) are omitted from the calculation of the \texttt{LOCFIT} fits and running medians for OMC-1/2/3/4/5. These inferred motion trends for the OMC-1/2/3/4/5 stars are further employed in the 3-D stellar kinematic movies described in \S \ref{sec_3dmovie}.

\begin{figure*}
	\includegraphics[angle=0.,width=150mm]{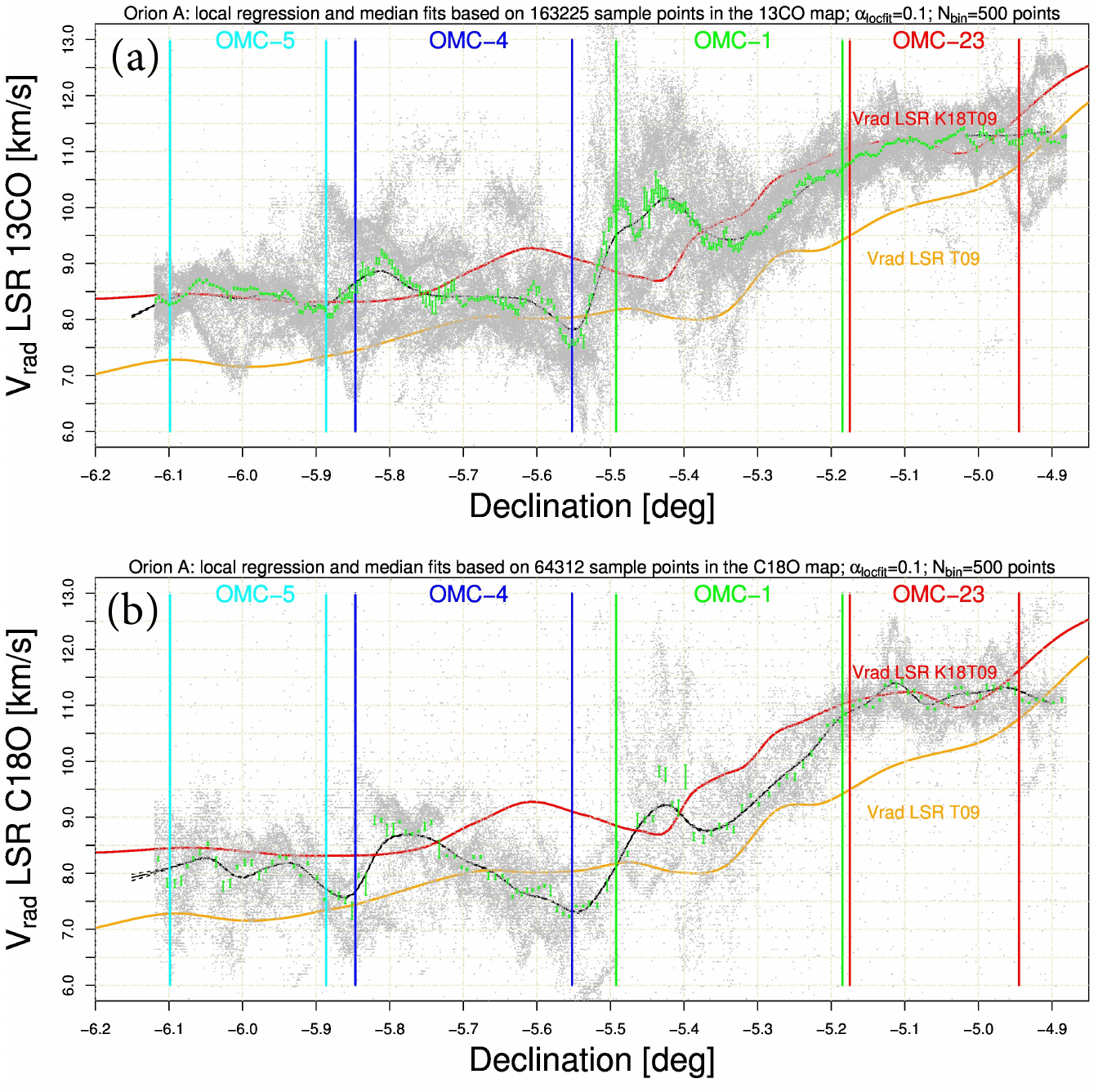}
	\caption{Gas position-velocity diagrams constructed by sampling the image pixels (gray) of the first-moment $^{13}$CO($1-0$) and C$^{18}$O$(1-0)$ emission maps from \citet{Kong2018}. The black solid and dashed curves indicate local quadratic regression fits and their $68$\% CIs, respectively, with the nearest neighbor bandwidth of $10$\%. The green points with error bars mark medians with their 68\% bootstrap uncertainties calculated from 500 pixels bins. The red curve presents the local regression fit to the stellar K18T09 radial velocities from Fig. \ref{fig_v_vs_declination}(c), and the orange curve indicates the local regression fit to the stellar radial velocities of \citet{Tobin2009}. }
	\label{fig_gas_ppvs}
\end{figure*}

Most of the Orion Head stars exhibit a smooth trend in velocity with range $\Delta V_X \sim 1$~km~s$^{-1}$, $\Delta V_Y \sim 1$~km~s$^{-1}$, and $\Delta V_Z \sim 3.5$~km~s$^{-1}$. This appears in all subsamples examined (\texttt{full} and \texttt{restricted}, ``K18T09'' and ``{\it Gaia} K18T09'' samples; see Supplementary material associated with  Figure~\ref{fig_v_vs_declination}).  

The $V_Z$ north-south trend is also seen in maps of molecular gas radial velocity in Orion A that ranges from $V_{rad,LSR} \ga 12$~km~s$^{-1}$ at the northern tip to $V_{rad,LSR} \la 2$~km~s$^{-1}$ at the southern tip of the entire cloud.  The gas motions are shown in Figure~\ref{fig_gas_ppvs}; similar figures appear in \citet{Furesz2008, Tobin2009, Hacar2016, DaRio2017, Kounkel2016}. Here \texttt{LOCFIT} and running median trends of stellar radial velocities for young stars are compared with molecular gas trends.  The datasets here are the ``K18T09'' sample of young stars from Table~\ref{tbl_vrad} (red curve) and the $^{13}$CO and C$^{18}$O first-moment emission maps from \citet{Kong2018} (black and green curves). The \texttt{LOCFIT} trend for the earlier stellar velocity dataset of \citet{Tobin2009} is also given for comparison (orange) showing an offset that is removed in APOGEE-derived velocities \citep{Kounkel2018}.

This empirical evidence of stellar-gas agreement in $V_Z$ leads to a simple but powerful result --- the bulk of young stars in Orion A still mimic the motion of their parental molecular material.  This was originally stated based on weaker datasets by \citet{Furesz2008}.  The major velocity discrepancies between gas and stellar radial velocities are around OMC-1 and OMC-4, perhaps due to the effects of OB star heating and protostar/star outflows. 

It is thus physically reasonable to define major stellar kinematic/spatial structures associated with the main molecular components seen in Figure~\ref{fig_v_vs_declination}: OMC-2/3, OMC-1, OMC-4, and OMC-5.  In addition to these major star structures, Figures~\ref{fig_smoothed_maps} and \ref{fig_v_vs_declination} show ``minor'' spatially clustered stellar components with peculiar kinematics: NGC~1977 main cluster, NGC~1977 south group, EONn and EONs groups, and Group~X (as well as two sparse groups of stars to the east and west of OMC-1/2/3/4 located outside the $83.6 < \alpha < 84\degr$ spatial stripe). Dynamical interpretation of the ``major'' and ``minor'' stellar components are presented in \S\S \ref{sec_ngc1977}-\ref{sec_discussion}.

Note that our selection of  kinematic stellar structures across the Orion A Head is conceptually different from those defined in \citet{Kounkel2018} who list 57 stellar groups over the area of the entire cloud, derived from their hierarchical clustering analysis of 6-dimensional data. Kounkel et al. caution that ``In many cases, the identified groups may not necessarily correspond to distinct subclusters... This is most apparent in the massive clusters: the ONC alone is associated with more than 30 groups.'' Since \citet{Kounkel2018} are interested in kinematic characterization of large-scale structures across the entire Orion Complex extending many degrees on the sky, $75<\alpha<95\degr$ and $-11<\delta<15\degr$, in their clustering analysis the authors choose to accept cluster structures with sizes up to $4\degr$.
	
Due to such a loose constraint their clustering procedure often results in stellar members of a single group being spatially distributed across multi-degree areas on the sky and members of multiple groups being spatially mixed. As a typical example, for instance, the members of their ``onc-7'' group (individually selected from their Table~2) are spread across the entire Head of Orion A showing loose spatial clustering features across NGC~1977, EONn, OMC-4, and south of OMC-5. Thus the kinematic properties inferred for individual stellar structures can not be directly compared between the Kounkel et al. and our current studies.

\subsection{{\it Gaia} Distance to the Head of Orion A}\label{sec_distance}

\begin{figure*}
	\includegraphics[angle=0.,width=170mm]{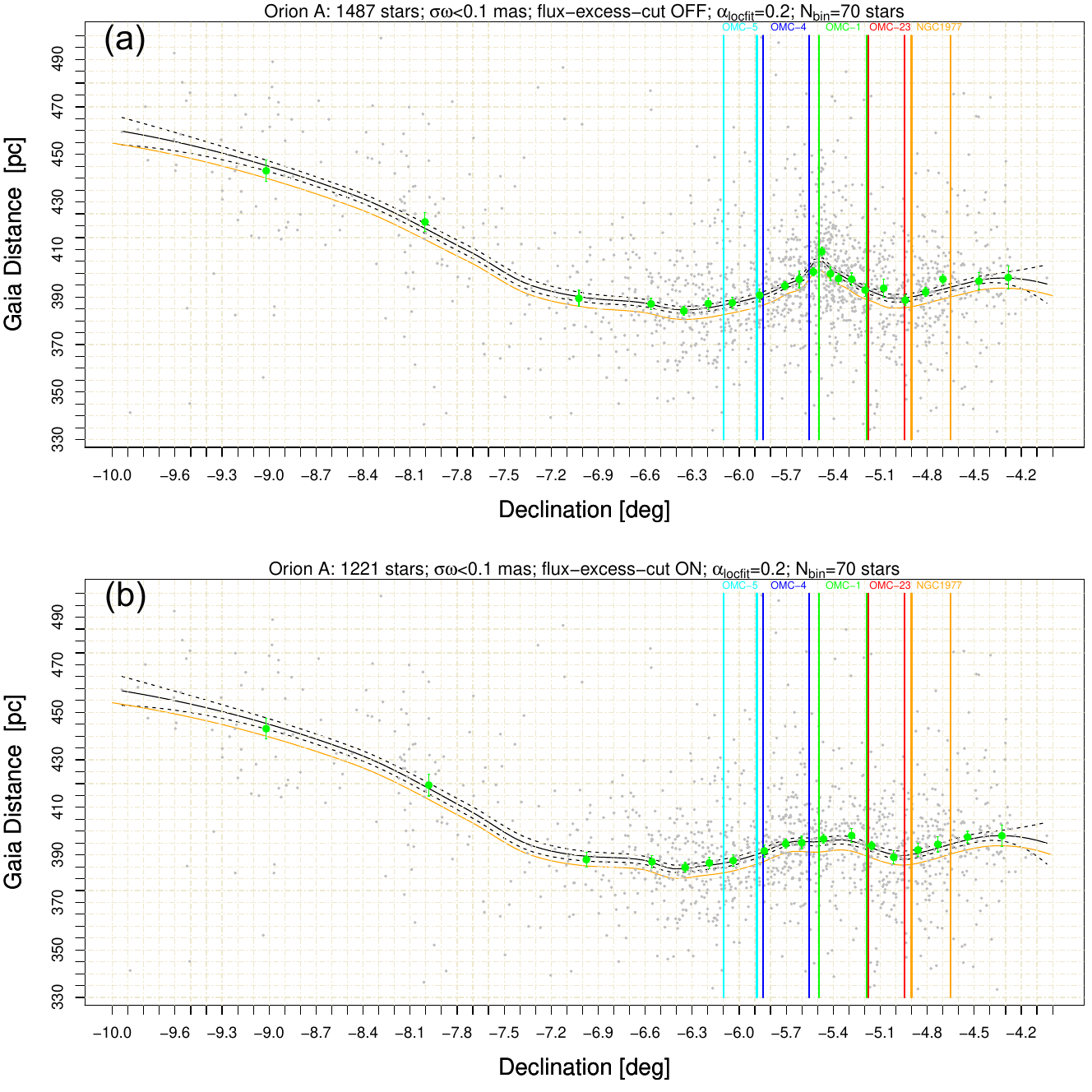}
	\caption{Distance $vs.$ declination for all {\it Gaia}-selected stars in the full Orion A region with $\sigma_{\bar{\omega}}<0.1$~mas for the \texttt{full} (panel a) and \texttt{restricted} (panel b) {\it Gaia} samples. The black solid and dashed curves indicate local quadratic regression fits and their $68$\% CIs with bandwidth of $20$\%. The green points with error bars mark medians with their 68\% bootstrap uncertainties calculated in adaptive bins with 70 stars. The orange curves indicate the local regression fits using the probabilistic distances of \citet{BailerJones2018}.}
	\label{fig_distance_vs_declination_oriona}
\end{figure*}

\begin{figure*}
	\includegraphics[angle=0.,width=180mm]{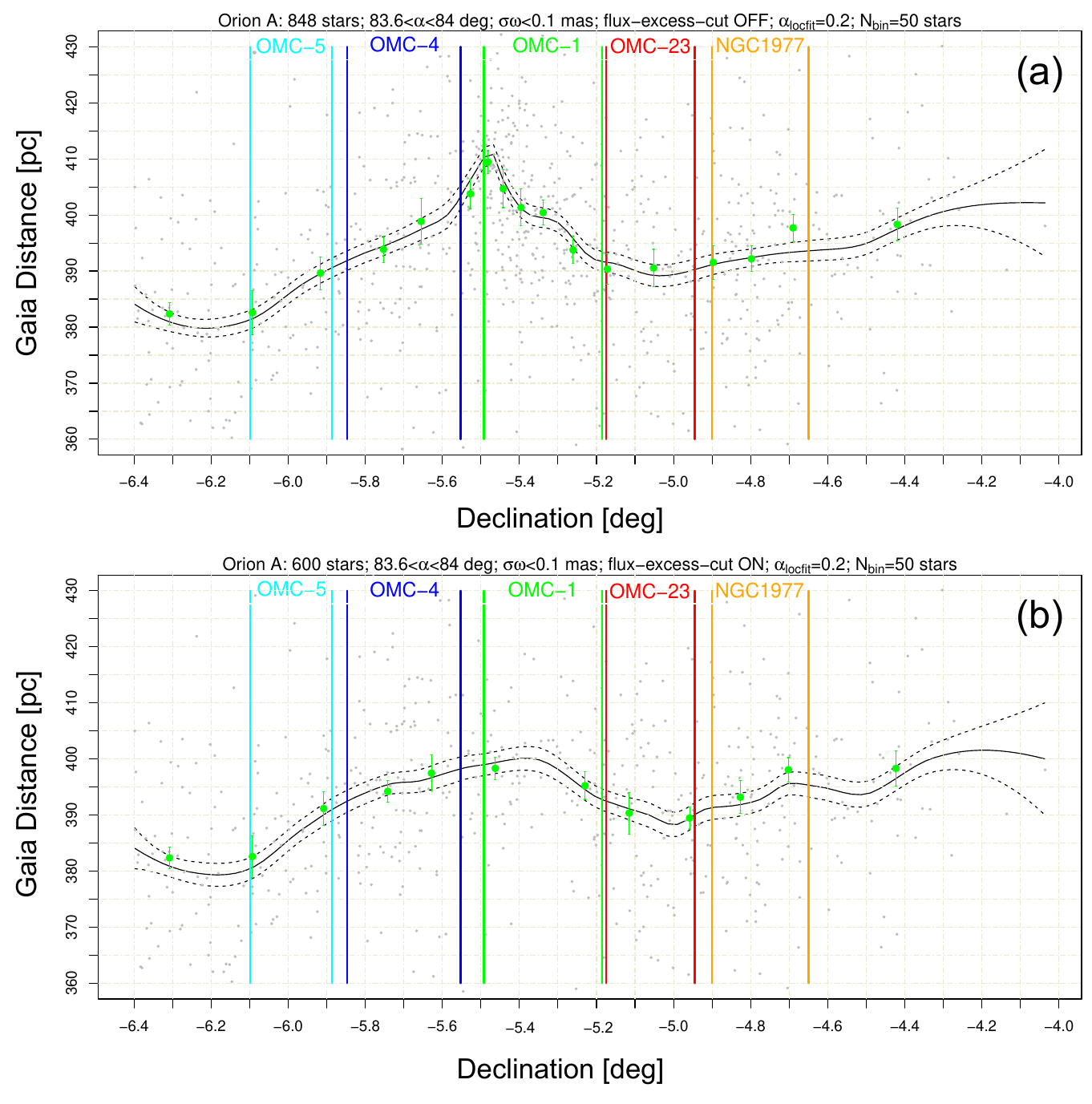}
	\caption{Inferred distance as a function of declination; same as Figure \ref{fig_distance_vs_declination_oriona} but restricted to the Head of Orion A within the $-6.4 < \delta < -4\degr$ and $83.6 < \alpha < 84\degr$ stripe. Panels (a) and (b) are for the \texttt{full} and \texttt{restricted} samples, respectively.}
	\label{fig_distance_vs_declination_head}
\end{figure*}

Figure \ref{fig_distance_vs_declination_oriona} shows inferred distances  based on {\it Gaia}  parallaxes as a function of declination across the entire Orion A cloud for both the \texttt{full} and \texttt{restricted} {\it Gaia} stellar subsamples. The plot confirms previous findings that the cloud's Tail is located further from the Sun than the cloud's Head, by $\sim 60$~pc \citep{Kounkel2018,Stutz2018b,Grossschedl2018a}.  We also support the findings of \citet{Kuhn2018} and \citet{Stutz2018b} showing that the stars in OMC-2/3 and OMC-5 are on average closer to the Sun than the stars in OMC-1 and OMC-4.  Star distance estimates based on \citet{BailerJones2018} (orange curve) retains the overall shapes of the distance distributions but shifts the distances towards lower values, by about 4~pc. The shape of the distance distribution across the Head (Figure \ref{fig_distance_vs_declination_head}) using our \texttt{full} {\it Gaia} sample is similar to that of Figure~3 in \citet{Stutz2018b}.  A slight bias arises due to {\it Gaia}'s inability to characterize heavily absorbed stars on the far side of the Orion A cloud.  The result is that distances towards disky stars are on average $\simeq 5$~pc further than the diskless stars (figure is not shown).


The results are surprising with high distance variations of up to $\Delta D \simeq 25$~pc across OMC-1/2/3/4/5 where the projected on the sky length of the Head is only $\sim 8$~pc; the cloud' Head appears to be oriented in the plane of the sky and is actually a sheet of gas and stars mostly oriented along the line-of-sight.  Independent lines of evidence support this interpretation. First, \citet{Stutz2018b} find that these distance trends strongly correlate with the trends of gas radial velocity across the Head of the cloud\footnote{They link this correlation to the presence of a standing wave in the cloud, while we argue below that is arises from the cloud's gravitational contraction.}. Second,  across the cloud's Tail that has a projected on the sky length of $\sim 26$~pc the variation in the distance to the Sun is about $\Delta D\sim 75$~pc \citep[][and Figure~\ref{fig_distance_vs_declination_oriona} here]{Grossschedl2018a}.  The Orion Tail thus has the same line-of-sight elongation as the cloud's Head.

As was mentioned before, due to possible distortions in PSF shapes, the {\it Gaia} parallaxes for stars with abnormally high BP/RP flux ratios in the \texttt{full} sample may be unreliable despite small formal statistical errors of $\sigma_{\bar{\omega}}<0.1$~mas. But it is puzzling to see that the distance distribution of these stars across OMC-1 is not random, which would be expected if their parallaxes were wrong. There is a clear spatial gradient in their inferred {\it Gaia} distances; stars with $D < 410$~pc preferentially reside in the central parts of OMC-1 where the gas velocity is below $10-11$~km~s$^{-1}$ (according to the $^{13}$CO gas velocity map of \citet{Kong2018}), whereas stars with $D > 410$~pc tend to lie near/in the areas with higher gas velocities of $\ga 11$~km~s$^{-1}$ (figure is not shown). The latter areas are associated with the expanding CO shells \#\# 10 (centered at $\alpha =83.93\degr$, $\delta=-5.46\degr$) and 11 (centered at $\alpha =83.79\degr$, $\delta=-5.46\degr$) from \citet{Feddersen2018}, presumably arising from excavation of molecular gas by UV radiation from OB stars and/or winds from nearby intermediate-mass stars. The positions of the stars located near/around the shell \# 10 are also projected against the low-extinction window of $A_V < 6$~mag from \citet{Lombardi2014}. 

These empirical facts allow us to speculate that if the {\it Gaia} parallaxes of the stars with high BP/RP flux excesses are accurate then many of these stars with $\delta \la -5.45\degr$ might be physically located behind the bulk of the {\it Gaia}-selected sample, perhaps within cavities protruding inside the cloud and/or on the other side of the cloud, and contributing to the apparent distance bump of $D>400$~pc at $-5.6<\delta<-5.4\degr$ seen in Figure \ref{fig_distance_vs_declination_head}a.

Even if the $-5.6<\delta<-5.4\degr$ distance bump of $D>400$~pc (Figure \ref{fig_distance_vs_declination_head}a) is ignored in our analysis, both star samples still indicate variable observer-to-cloud distances towards different parts of the cloud. For instance, the distance variation across the OMC-1 alone could be of a few-to-several parsecs (Figure \ref{fig_distance_vs_declination_head}b). Along with consideration of the aforementioned systematic distance differences of $\sim 5$~pc between disky and diskless stars and similarity in the line-of-sight elongations between the cloud's Tail and Head, it is thus reasonable to suggest that the Head is ``V''-shaped (or has a horse-shoe morphology) along the line-of-sight (see 3-D kinematic movies in Section \ref{sec_3dmovie}). These facts allow us to speculate that the central part of ONC, a densely concentrated cluster in front of the OMC-1 cloud with a projected core radius of $\sim 0.2$ parsecs and an extent of $\sim 2$~pc \citep{Hillenbrand1998}, may have an elongated (a few to 5~pc) ``sausage''-like morphology along the line-of-sight. 

For {\it Gaia} stars, Figures~\ref{fig_age_maps}(b,c) show raw and adaptively smoothed maps of ages from \citet{DaRio2016}. It is interesting to note that the stars associated with the southern part of OMC-5 and Group-X are generally older than the OMC-1/4 stars. This fact allows us to speculate that some stars associated here with OMC-5 could be members of an older, foreground stellar cluster, NGC~1980 \citep{Alves2012}. This possible ``contamination'' by NGC~1980 stars might lead to underestimation of cloud distances, especially towards the OMC-5 part of the filament.
 
\subsection{Stellar Groups with Discrepant Kinematics:  NGC 1977, Extended Orion Nebula, and Group~X}\label{sec_ngc1977}

Located at the northern end of the Orion A cloud, NGC 1977 with the neighboring NGC 1973 and NGC 1975 HII regions (also collectively known as OB association Ori OBc subgroup 2) is a moderately rich stellar cluster residing within an HII bubble of complex structure \citep{Peterson2008}.  The region is ionized by a B1V star HD~37018, B2V star HD~37058 = 42 Ori, and B3V stars HD~294264 and HD~36958 \citep{Skiff2009}.   

Located $\sim 30^\prime$ north of the Orion Nebula, it is seen with the naked eye  as the northern `star' in the sword of Orion.  It was first imaged by \citet{Ricco1895} with spectroscopic characterization by \citet{Hubble1922} and molecular study by \citet{Kutner1976}.  The stellar content of NGC 1977 has a high concentration of several dozen H$\alpha$-emitting T Tauri stars \citep{Gomez1998}, 170 stars with infrared-excess and/or variability \citep{Peterson2008}, two embedded luminous protostars \citep{Mookerjea2000}, 7  cometary proplyds \citep{Kim2016}, and 260 X-ray/IR emitting young stars \citep{Getman2018b}. The median age for the {\it Gaia} stars in the cluster is about 2~Myr ($Age_{D16} = 2.1 \pm 0.2$~Myr; $Age_{JX} = 1.9 \pm 0.3$~Myr; Table \ref{tbl_other_props}).  The median age for the X-ray and IR-selected young stars that are not part of our {\it Gaia} sample could be even lower, $Age_{JX} = 1.7 \pm 0.2$~Myr with an inferred cluster disk fraction of $0.43 \pm 0.07$ \citep{Getman2018b,Richert2018}.

Cluster stars and the gas that has not yet been expelled have relatively high radial velocities of $V_{rad,LSR} > 10$~km~s$^{-1}$ \citep{Tobin2009, DaRio2017}, at the highest end of the north-south star-gas velocity gradient (\S\ref{sec_stellargroups}). Tobin et al. stress that the stars in this cluster have a relatively narrow $V_{rad}$ distribution compared to the stars in Orion A and suggest that many high-velocity stars have already fled the region. Figure \ref{fig_smoothed_maps} also shows the cluster has a uniquely high $V_Y$ velocity compared to the bulk of the stars in OMC-1/2/3/4.

For the comparative analysis below, to increase counting statistics, we combine the stars in the NGC~1977 main cluster with the stars in the NGC~1977 south group, since both have roughly similar kinematics and distances judging from Figures \ref{fig_smoothed_maps} and \ref{fig_v_vs_declination}. The stars in the southern group may be slightly older ($Age_{D16} = 2.8 \pm 0.7$~Myr; Table \ref{tbl_other_props}) and are also located in the region mainly devoid of molecular gas (Figure \ref{fig_lombardi_map}).

The Extended Orion Nebula (EON) cavity southwest of the Orion Nebula is a large bubble filled with plasma emitting soft X-rays  \citep{Gudel2008}.  Two stellar groups with peculiar kinematics are projected against the EON, designated EON-north (EONn) and EON-south (EONs) in Figures \ref{fig_lombardi_map} and \ref{fig_smoothed_maps}. The EONn group was originally noticed by \citet[][their Figure  9]{Furesz2008} for its unusually high number of stars with blue-shifted radial velocities, compared to the nearby stars and gas. They proposed that these stars were formed in molecular structures with originally blue-shifted velocities and the gas from these structures was efficiently expelled by nearby outflows. From our preliminary examination of Figures~\ref{fig_smoothed_maps} and \ref{fig_v_vs_declination} and the data in Table~\ref{tbl_other_props}, we find that both EONn and newly identified EONs groups have similar kinematics, distances, and age properties. To increase counting statistics, for the comparative analysis below, the two groups are merged together.

\begin{figure*}
 	\includegraphics[angle=0.,width=180mm]{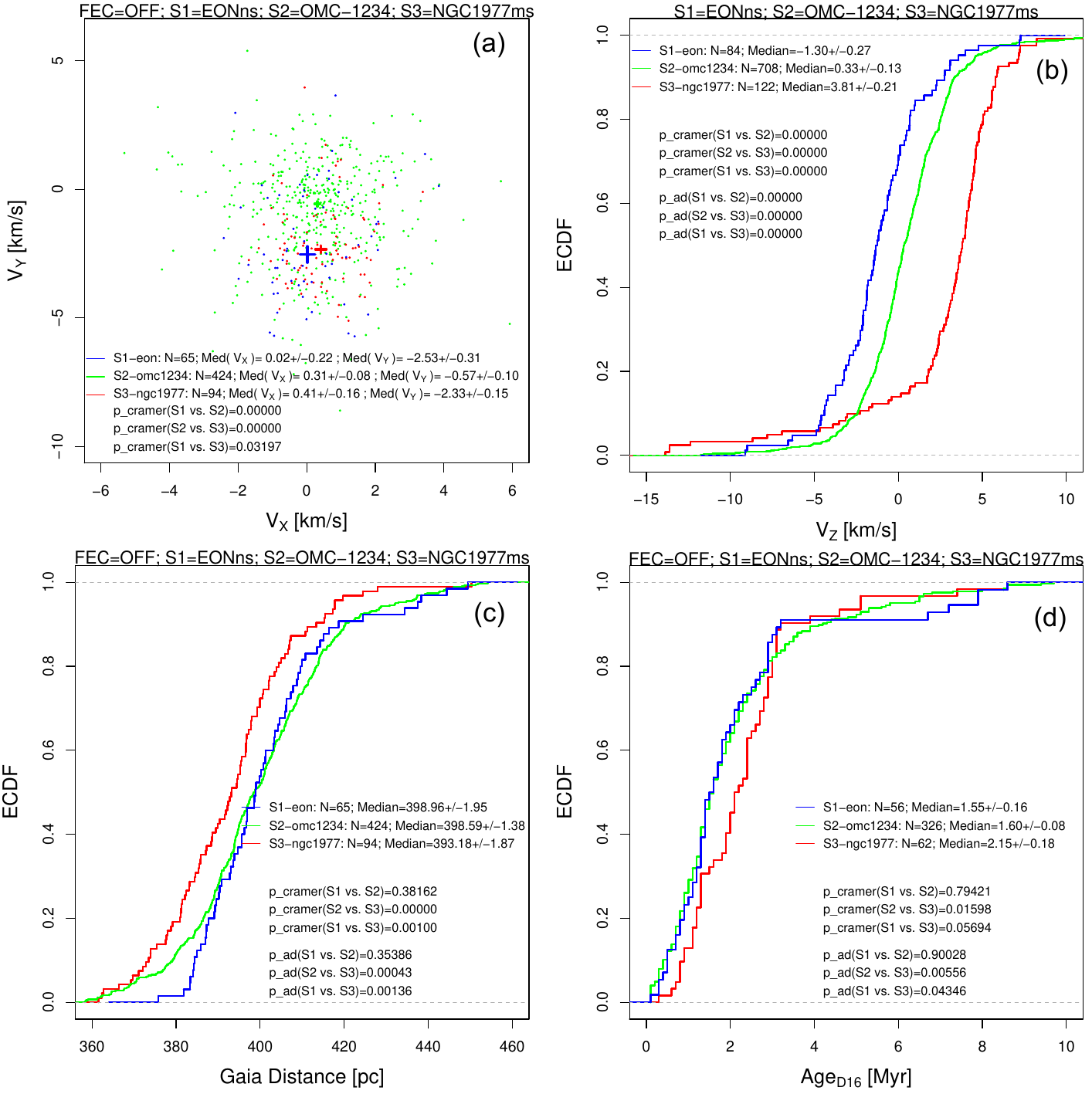}
 	\caption{Comparison of stellar kinematics, distances, and ages for three star samples that reside within the $-5.8 < \delta < -4.6\degr$ and $83.6 < \alpha < 84\degr$ stripe encompassing OMC-1/2/3/4: S1-eon (blue) includes stellar members of the EONn and EONs groups; S3-ngc1977 (red) represents members of the NGC~1977m cluster and NGC~1977s group; and S2-omc1234 (green) encompasses stars that lie projected against OMC-1/2/3/4.  (a) $V_Y - V_X$ transverse velocity diagram with Medians and 68\% CIs for the three stellar groups, shown as crosses.  The panel legends give the numbers of stars, median velocities and their 68\% bootstrap errors, and $p$-values for the two-sample non-parametric tests for equality. (b,c, and d) Empirical cumulative distribution functions (ECDFs) of $V_Z$, {\it Gaia} distance, and age for the three stellar samples. The diagrams (a,c, and d) are presented for the {\it Gaia} \texttt{full} star sample. The $V_Z$ diagram is given for the ``K18T09'' star sample. Similar diagrams for the {\it Gaia} \texttt{restricted} sample are provided in the Supplementary Materials. The Supplementary figure also includes a $V_Z$ panel for the ``{\it Gaia}-K18T09'' star sample. The format of this figure is repeated in the following two figures.}
 	\label{fig_ngc1977eon_kinematics}
 \end{figure*}

Figure \ref{fig_ngc1977eon_kinematics} shows comparison of kinematics, distances, and ages among the EONn$+$EONs (S1-eon), OMC-1/2/3/4 (S2-omc1234), and NGC~1977 main$+$south (S3-ngc1977) star samples. All the three samples have statistically significantly different kinematics. With respect to the bulk of the stars in OMC-1/2/3/4, the S1-eon and S3-ngc1977 stars have significant $V_Y$ motions towards south ($\Delta V_Y \sim 2$~km~s$^{-1}$) and $V_Z$ motions towards (S1-eon; $\Delta V_Z \sim 1.5$~km~s$^{-1}$) and away from (S3-ngc1977; $\Delta V_Z \sim 3.5$~km~s$^{-1}$) the observer. The EON stars seem to be younger than the NGC~1977 stars; this supports the idea of \citet{Furesz2008} that the removal of the EON-related parental gas was quite rapid. The \texttt{full} {\it Gaia} sample suggests that the NGC~1977 structures may be located slightly (by a few parsecs) closer to the observer than the bulk of the OMC-1/2/3/4/EON stars; however this is not the case for the \texttt{restricted} sample, for which all the three stellar samples have similar distances. 

\begin{figure*}
 	\includegraphics[angle=0.,width=180mm]{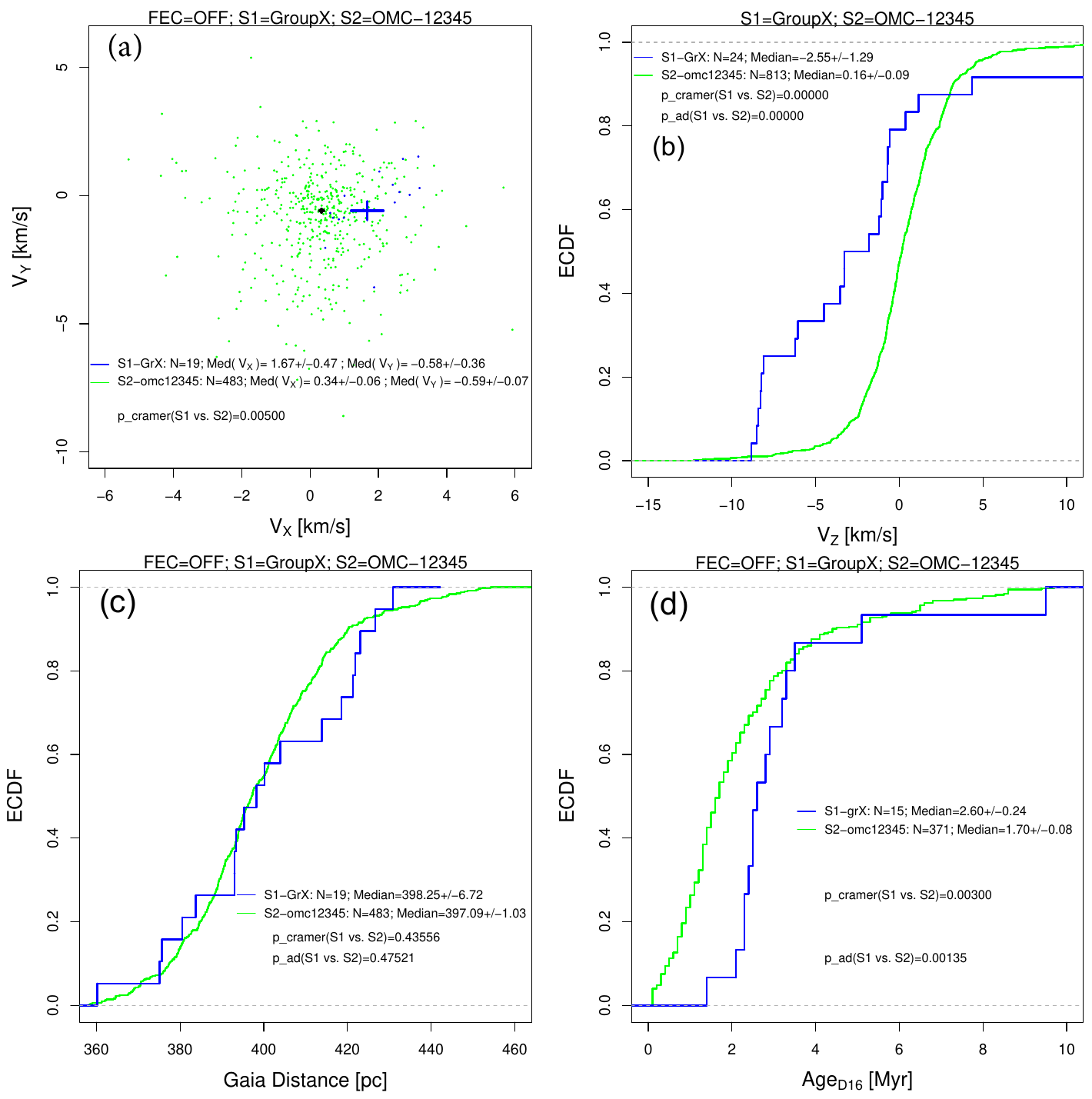}
 	\caption{Comparison of stellar kinematics (a,b), distance (c), and ages (d) for two star samples: Group-X (S1; blue) and OMC-1/2/3/4/5 (S2; green).}
 	\label{fig_groupx_kinematics}
 \end{figure*}

Finally, we note a prominent stellar group with peculiar kinematics designated Group~X comprised of $\sim 20$ pre-main sequence stars located about $3$~pc to the west of OMC-5. The group lies projected against a few "bullet"-like dusty structures elongated along the south-west direction (Figure~\ref{fig_lombardi_map}) but it is not clear whether these stars and gas structures are kinematically linked. Figure \ref{fig_groupx_kinematics} shows that relative to the OMC-1/2/3/4/5 stars, Group~X has significant motions along the X- and Z-directions of $\Delta V_X \ga 1$~km~s$^{-1}$ and $\Delta V_Z \ga 2.5$~km~s$^{-1}$ towards west and towards the observer, respectively. This group is relatively old, $Age_{D16} = 2.6 \pm 0.2$~Myr. Its median distance of $398 \pm 7$~pc is indistinguishable from the full OMC-1/2/3/4/5 sample. 

Thereby, in addition to the earlier findings on relatively high radial velocities in NGC~1977-main and EONn \citep{Furesz2008,Tobin2009} we present new empirical evidence for uniquely, relatively high $V_Y$ and $V_Z$ velocities of the NGC~1977 main, NGC~1977 south, EONn, and EONs stars compared to the bulk of the stars in OMC-1/2/3/4. We suggest that the NGC~1977 and EON groups are caught in a dynamic state of departure from their birthplaces. Group~X is perhaps an example of a surviving bound stellar group that has recently experienced such a departure.

\subsection{Drifting of Stars East and West of OMC-1/2/3/4}\label{sec_eastweststars}

\begin{figure*}
 	\includegraphics[angle=0.,width=180mm]{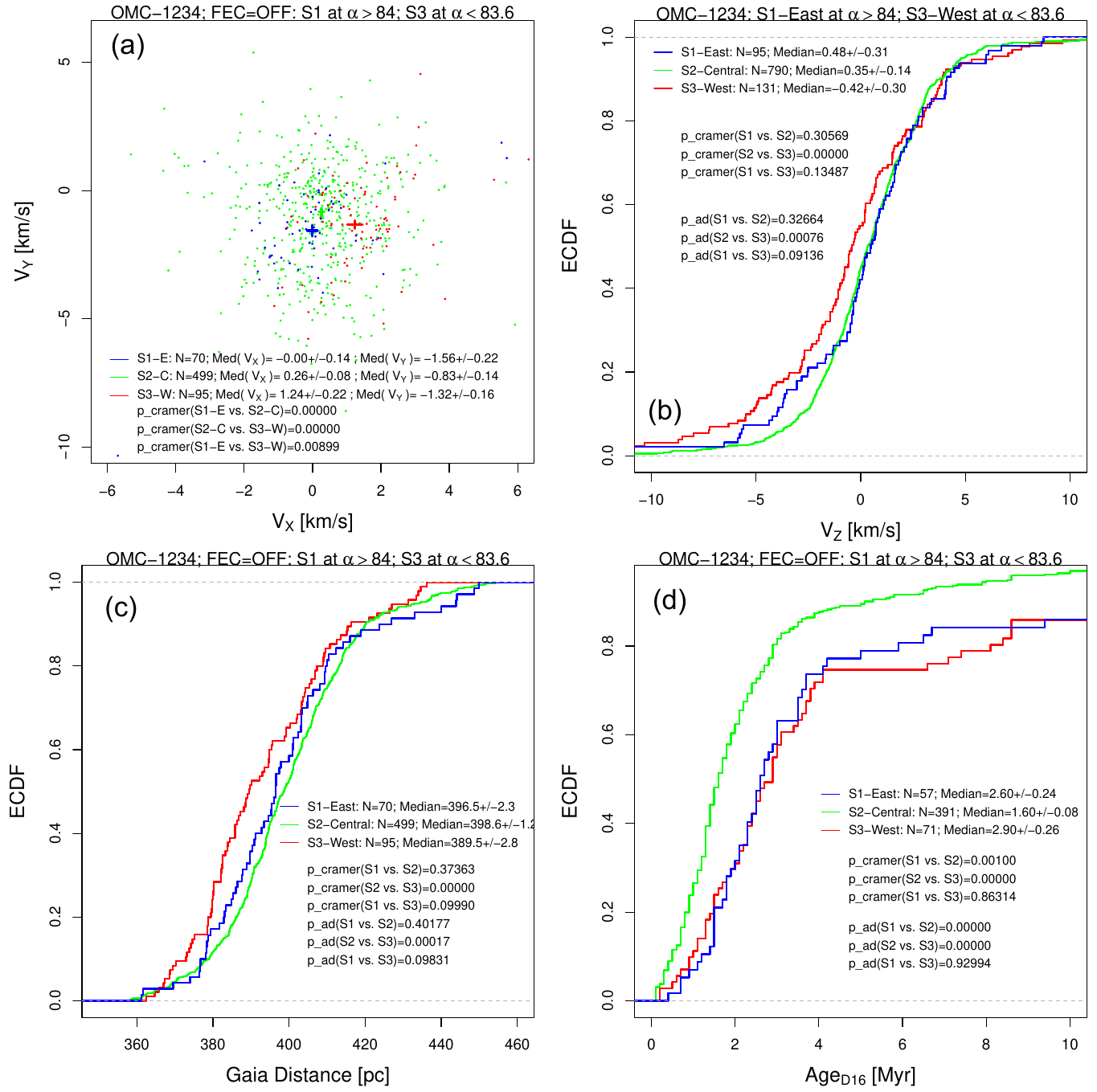}
 	\caption{Comparison of stellar kinematics, distances, and ages for three star samples that reside within the $-5.8 < \delta < -4.9\degr$ stripe encompassing OMC-1/2/3/4: S1-east (blue) and S3-west (red) are located to the east and west of OMC-1/2/3/4, respectively; and S2-central (green) lies projected against OMC-1/2/3/4.}
 	\label{fig_westeast_kinematics}
 \end{figure*}

Within the $-5.8<\delta<-4.9\degr$ spatial stripe, there is a noticeable number of young stars with $\alpha > 84\degr$ (denoted here as S1-East sample) and with $\alpha<83.6\degr$ (denoted as S3-West sample) whose $V_X$ velocity components indicate motions away from the cloud (Figure~\ref{fig_smoothed_maps}b). We compare kinematics, distances, and ages of these stars to the OMC-1/2/3/4 stars, denoted here as the S2-Central sample in Figure \ref{fig_westeast_kinematics}.  The $V_X$ and $V_Y$ motions and stellar ages are significantly different between the OMC-1/2/3/4 stars and either the S1-East or S3-West samples. The S1-East and S3-West stars are older than the OMC-1/2/3/4 stars.  Unlike the S1-East stars, the S3-West stars have significant Z-motions away from OMC-1/2/3/4, towards the observer, that are consistent with them being somewhat closer to the observer than the OMC-1/2/3/4 stars. 

 Overall, these results suggest that the S1-East and S3-West stars were born in or near the location of the OMC-1/2/3/4 cloud, roughly 2.7 Myr ago.  They are drifting away either from the rich ONC or from the distributed star formation along the molecular filament at a speed around 1~pc~Myr$^{-1}$.  For S3-West, this drifting star picture is consistent with the $V_Z$ measurements and {\it Gaia} distance estimates (in the case of the \texttt{restricted} sample). However, a distance difference of several parsecs is seen between the OMC-1/2/3/4 and S3-West in the case of the \texttt{full} sample suggesting it may have formed in some not-dissipated cloudlet in front of OMC-1/2/3/4 roughly 3~Myr ago.
 
 \clearpage
 \newpage
 \clearpage
 \newpage
 
 \clearpage
 \newpage
 
 \clearpage
\newpage

\subsection{3-D Stellar Kinematics}\label{sec_3dmovie}
\begin{figure*}
	\includegraphics[angle=0.,width=170mm]{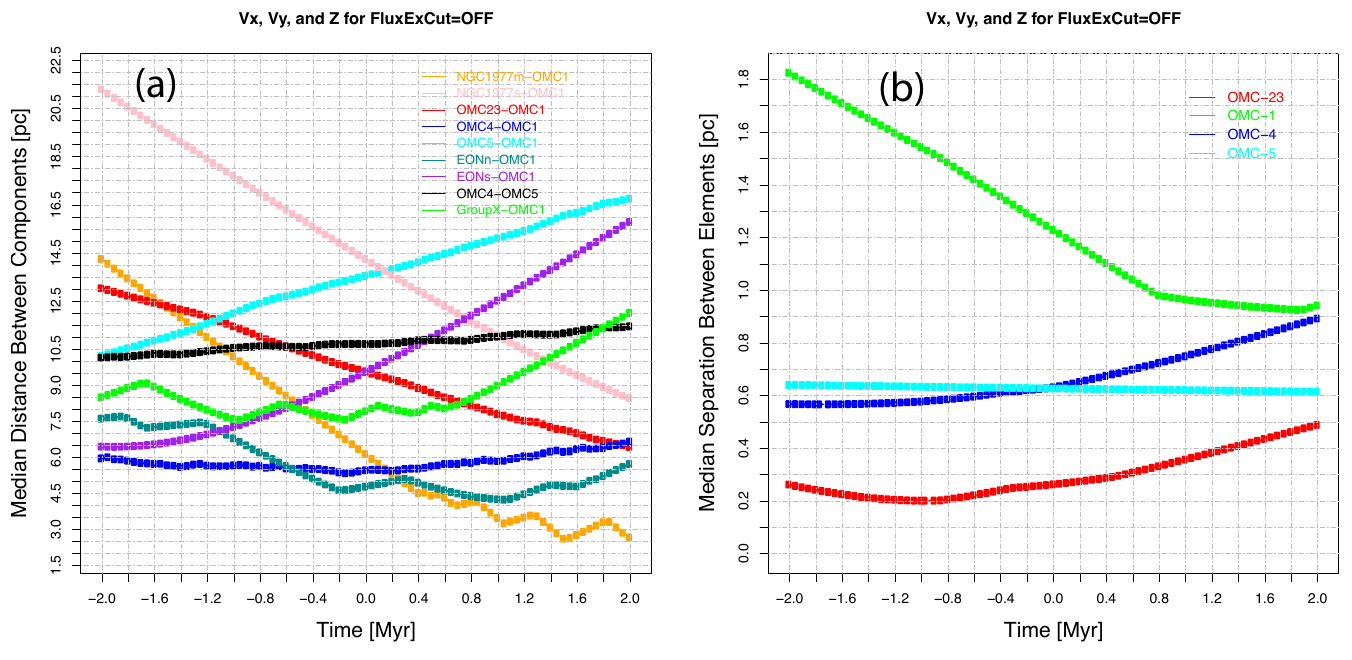}
	\caption{For the  {\it Gaia} \texttt{full} sample, two-dimensional projections from the 3-D kinematic movies showing changing distances between main stellar structures (a) and changing separations between neighboring elements within main stellar structures (b). Similar diagrams for the {\it Gaia} \texttt{restricted} sample are provided in the Supplementary Materials.}
	\label{fig_temporal_evolution_plots}
\end{figure*}

\begin{figure*}
	\includegraphics[angle=0.,width=170mm]{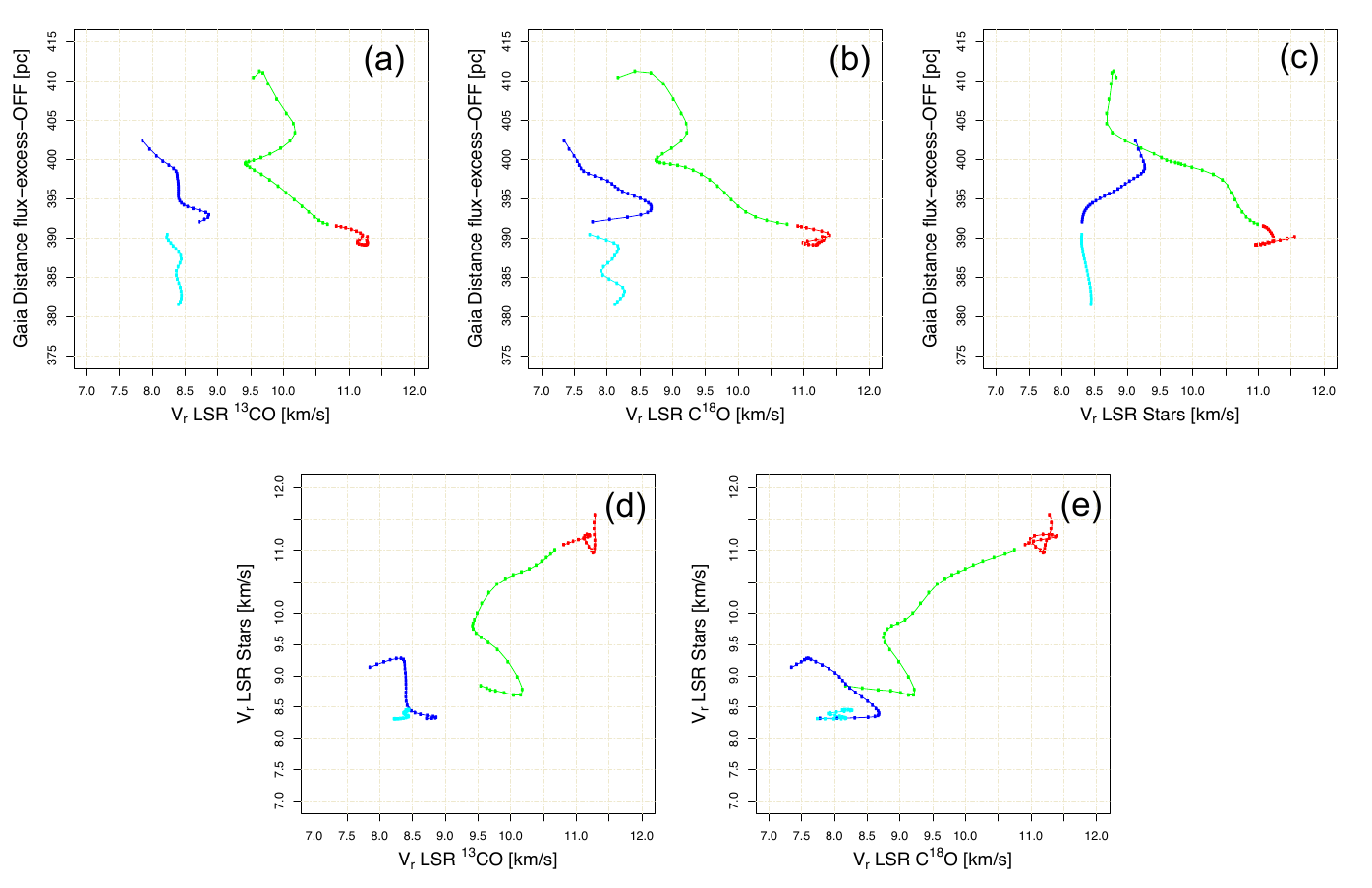}
	\caption{For the Head of Orion A within the $83.6 < \alpha < 84\degr$ stripe, comparison of the local regression fits for the {\it Gaia} distances (from Fig. \ref{fig_distance_vs_declination_head}), gas radial velocities (from Fig. \ref{fig_gas_ppvs}), and stellar radial velocities (from Fig. \ref{fig_v_vs_declination}c). Panels (a, b, c) show  distance-velocity trends. Panels (d and e) give star-gas velocity trends. Points associated with different cloud/star structures are color-coded as: OMC-2/3 (red), OMC-1 (green), OMC-4 (blue), and OMC-5 (cyan). {\it Gaia} distances are given for the \texttt{full} star sample. Diagrams with distances for the \texttt{restricted} sample are provided in the Supplementary Materials.}
	\label{fig_distance_velocity_trends}
\end{figure*}

 The Supplementary Materials present movies depicting 3-D kinematics of main stellar components in the Head of Orion A, assuming constant velocities, starting 2~Myr in the past and continuing 2~Myr into the future. Nine stellar structures have independent motions in the movies: OMC-2/3, OMC-1, OMC-4, OMC-5, NGC~1977main, NGC~1977south, EONn, EONs, and Group~X. The X, Y, Z, $V_X$, $V_Y$, $V_Z$, and {\it Gaia} distance ($D$) quantities are taken from Figures \ref{fig_smoothed_maps}, \ref{fig_v_vs_declination}, \ref{fig_distance_vs_declination_head},  \ref{fig_ngc1977eon_kinematics},  \ref{fig_groupx_kinematics}, and \ref{fig_westeast_kinematics}. The NGC~1977main, NGC~1977south, EONn, EONs, and Group~X groups are represented as single points while the OMC-2/3, OMC-1, OMC-4, and OMC-5 structures are represented each by 16 points (following the velocity \texttt{LOCFIT} trends) uniformly sampled along the $\alpha-\delta$ \texttt{LOCFIT} trend of the {\it Gaia} stars in OMC-1/2/3/4/5. This stellar backbone runs nearly parallel but $\sim 3\arcmin$ to the east of the OMC-1/2/3/4 C$^{18}$O backbone but passes through the middle of the OMC-5 C$^{18}$O structure. Two versions of the movie are constructed corresponding with \texttt{full} $vs.$ \texttt{restricted} samples. 

Figure \ref{fig_temporal_evolution_plots} summarizes main results emerged from the movies in two 2-D projections: the evolution of median distances between OMC-1 and the remaining stellar components; and the evolution of median separations between neighboring elements (16 points) for the 4 main stellar structures (OMC-1, OMC-2/3, OMC-4, and OMC-5). Plots for the \texttt{full} permutation are presented here; similar plots for the \texttt{restricted} permutation are provided in the Supplementary Materials.

Three results emerging from the movies reveal the complexity of motions of star groups produced in a giant molecular cloud:
\begin{enumerate}

\item The stars in NGC~1977 are departing from their birthplaces and are on courses of approaching OMC-1, while the stars in EON (Group~X) are departing (have already departed) their birthplaces and are moving away from OMC-1. 

\item Stars related to OMC-1 are in the state of a 3-D contraction, OMC-2/3 and OMC-4 are in a state of expansion, and the stars in OMC-5 are static with respect to each other. 

\item At present, the stars in OMC-2/3  are moving towards OMC-1, OMC-5 is moving away from OMC-1, and OMC-4 show no motion with respect to OMC-1. This means that the stars in the northern part of the Head (OMC-2/3 and OMC-1) show a tendency for a global contraction  while the stars in the southern part of the Head (OMC-4 and OMC-5) show a tendency for a global expansion.
\end{enumerate}

\subsection{Relationship of Gas and Stars}\label{sec_relationship_d_vs_vr}

Figure \ref{fig_distance_velocity_trends} shows comparison among the \texttt{LOCFIT} trends of {\it Gaia} distances (from Figure~\ref{fig_distance_vs_declination_head}), gas radial velocities (from Figure \ref{fig_gas_ppvs}), and stellar radial velocities (from Figure \ref{fig_v_vs_declination}). The trends are sampled at every $\Delta \delta = 0.01\degr$ point.


Panels a and b show that across each of the individual molecular cloud components (OMC-1, OMC-4, OMC-5, and the southern part of OMC-2/3) the gas velocities anti-correlate with the {\it Gaia} distances.  That is, more distant parts of the cloud move slower from the observer than closer parts of the cloud.  A similar plot in \citet{Stutz2018b} supports their scenario that a standing wave is present across the Head of the cloud. In \S \ref{sec_discussion} we provide an alternative explanation of these correlations that involve gas gravitational contraction.

Panel c of Figure  \ref{fig_distance_velocity_trends} shows that the stellar radial velocities  anti-correlate with the {\it Gaia} distances in OMC-1 and OMC-5, correlate in OMC-4, and show no relation in OMC-2/3. Panels d and e indicate positive gas-star velocity correlations in OMC1 and OMC-5,  anti-correlation in OMC-4, and no effect in OMC-2/3. 

These results suggest that the bulk of the {\it Gaia} stars in OMC-1 and OMC-5 are in a dynamical state of contraction along the line of sight, while OMC-4 is expanding and OMC-2/3 is static.

\section{Discussion}\label{sec_discussion}

In this study, we present the structure and kinematics of stellar groupings in the Head of the Orion A cloud, in particular reporting groups exhibiting peculiar kinematics (\S\S \ref{sec_ngc1977} and \ref{sec_eastweststars}). The cluster NGC~1977 and stellar groups NGC~1977 south, EON-north, and EON-south are caught in the act of departing their birthplaces. The older Group~X and S3-West stellar groups have already likely moved away from the places of their origin. These kinematic results are confirmed independently using data from both ground-based ($V_Z$ component) and {\it Gaia} telescopes ($V_X$ and $V_Y$).  

Our work is based on a large sample of stars derived from X-ray, infrared excess, and optical spectroscopic indicators of stellar youth.  These samples of pre-main sequence stars are derived from carefully designed studies using the Chandra X-ray Observatory, Spitzer Space Telescope, Hubble Space Telescope, and ground-based telescopes.   Little contamination by field stars is likely to be present \citep{Feigelson2013}.  Employed here radial velocities were previously obtained with the Sloan APOGEE spectrograph, Hectochelle on the MMT, and MIKE Fibers on the Magellan Clay telescope, and care is taken to include only reliable kinematic and distance information from {\it Gaia}. With a large uncontaminated sample, our study can uncover detailed patterns that were not adequately revealed in previous work. 

We confirm earlier findings that, although the molecular filament seems to lie in the plane of the sky along a north-south axis, different parts of the cloud have considerably different distances from the Sun.  We then confirm that the {\it Gaia} distances and kinematics are reflected in different gas velocities, indicating a close relationship between stars and the surrounding gas in parts of the cloud. Our 3-D kinematic analyses further demonstrate contraction and expansion of stars associated with different parts of the cloud.

Our major distance/kinematic results appear qualitatively similar for all subsamples examined (\texttt{full} vs. \texttt{restricted}, ``K18T09'' vs. ``{\it Gaia} K18T09''). 

Overall, our work and related studies provide a detailed view of the structure and kinematics of stars as the outcome of star formation in a giant molecular cloud.  The situation is complex: no simple scenario of gravitational collapse, filament infall or instabilities, colliding clouds, or propagating star formation can explain the breadth of phenomena, such as the twisting motions seen in our movies.  We now discuss explanatory scenarios in more detail.

\subsection{Scenarios for the Unusual Motion of NGC~1977} \label{sec_ngc1977_head_discussion}

NGC~1977 lies at the northern end of a molecular filament that exhibits a strong large-scale velocity gradient.   It has $V_X$ and $V_Y$ velocity components that differ by about 3 km s$^{-1}$ from neighboring stars associated with the backbone of the cloud, and is thus separating from the cloud by $\sim 3$ pc Myr$^{-1}$.  Five independent explanations can be considered. Among those, the ``hierarchically collapsing cloud'' and ``feedback by OB stars'' scenarios are more consistent with our data and/or prone to less problems and limitations than the ``turbulent clouds'', ``cloud-cloud collision'', and ``slingshot'' scenarios (see below). This does not imply exclusion of the latter three models from consideration. For instance, while turbulence in giant molecular clouds became less important as previously thought, the twisting motions of the Orion Head (seen in our 3-D movies) would be consistent with the presence of gas eddies curving in different directions as expected from the ``turbulent clouds'' scenario given below.

{\bf Hierarchically collapsing cloud} ~~  Sections \ref{sec_distance}, \ref{sec_3dmovie}, and \ref{sec_relationship_d_vs_vr} show velocities that are anti-correlated with distances in different parts of cloud's Head. This was previously reported by \citet{Stutz2018b}.  A simple explanation involves contraction within each of the cloud components (OMC-1, OMC-2/3, OMC-4, and OMC-5)\footnote{Notice that our recent {\it Gaia} study by \citet{Kuhn2018} reports some evidence for mild dynamical expansion of stars in the central part of the ONC cluster that lies within the COUP field. However, \citet{Kuhn2018} consider only 2-D kinematics with no involvement of radial velocities. $V_Z$ is the dominant velocity component in OMC-1 (\S \ref{sec_stellargroups}); and it is primarily responsible for the star contraction in OMC-1 presented in our 3-D movies using permutations with variable distances.}, and between cloud components.  This interpretation agrees with models of gravitationally collapsing large molecular clouds \citep{Colin2013,Vazquez-Semadeni2017,Hartmann2007,Kuznetsova2018,Vazquez-Semadeni2019} that involve infall motions (sometimes called  `gravitational focusing') of gas and forming stellar subclusters near the center of the gravitational potential. We recall that the hierarchical collapse models of \citet{Colin2013}, \citet{Vazquez-Semadeni2017}, and \citet{Vazquez-Semadeni2019} are also promising in explaining observed core-halo age gradients in numerous MYStIX$+$SFiNCs stellar clusters, including the ONC \citep[][see also Appendix \S \ref{appendix_b} here]{Getman2014b} and other clusters \citep{Getman2018a}.

Infall of stellar components is also consistent with the observations of gas infall in OMC-1 \citep{Hacar2017} and OMC-1/2/3/4 \citep{Wu2018}. For example, Hacar et al. interpret a V-shaped profile in the position-velocity distribution of IRAM N$_2$H$^{+}$ emission in the cloud's Head as a signature of gas gravitational collapse in OMC-1.  Wu et al. find variations in relative velocities of different molecular tracers (NH$_3$, $^{12}$CO, $^{13}$CO, and N$_2$H$^{+}$) that they interpret as gas accretion flows within each of the components.

The simultaneous presence of gas and stellar contraction is expected from star formation simulations.  Theoretical models featuring gas infall towards gravitational potential predict dynamical stellar subcluster mergers \citep{Maschberger2010,Bate2012,Vazquez-Semadeni2017,Kuznetsova2018,Vazquez-Semadeni2019}.   In particular, the converging motions of OMC-2/3 and OMC-1 stars, clearly seen in the 3-D movies, could lead to a subcluster merger in the near future. Having a relatively high $V_Y$ component, the NGC 1977 cluster itself is approaching OMC-1 (see the movies in Section \ref{sec_3dmovie}); this fact may also support the scenario of a gravitationally contracting filament with coalescing stellar sub-structures. We note, however, that {\it Gaia} stellar motions in numerous MYStIX$+$SFiNCs regions with multiple subclusters provide no signs of such mergers  \citep{Kuhn2018}.  This suggests that the coalescence of subclusters is rapid, occurring when most stars are still embedded and not available for {\it Gaia} to see in most giant molecular clouds.  We see this event in the Orion cloud only due to its proximity and the size of the stellar sample available for study. 

The twisting motions of the cloud's Head seen in our movies may be compatible with the  rotation-induced twisting and/or collapse motions enhancing anisotropies expected from the ``collapsing cloud'' scenario \citep{Hartmann2007, Vazquez-Semadeni2019}.

{\bf Feedback by OB stars} ~~  The NGC~1977 region lies at the edge of one of a large infrared bubbles centered on the location of Ori OB1ab stars.  This can be seen in 4$\degr$ radius WISE~12~$\mu$m image centered at $(\alpha, \delta) = (84.0\degr,-2.5\degr)$.  This location suggests that older generations of massive Ori O1ab stars may have an external influence on the kinematics and morphology of Orion A cloud \citep{Grossschedl2018a}, including the unusual motion of NGC~1977. \citet{Bally1987} proposed that the compressed shape and large-scale velocity gradient of the Orion Head cloud are caused by compression of an original cloud and deposition of additional interstellar material by HII region expansion, stellar winds, and supernova ejecta from hundreds of early-generation OB stars. Similarly, based on their large scale $^{12}$CO survey data for the Orion-Monoceros complex, \citet{Wilson2005} argue that the gas in the proto-Orion A cloud was compressed by the effects of Ori OB 1b stars, perhaps triggering star formation in Orion A. 

Related to these scenarios is the recent discovery of two older ($t \ga 4-7$~Myr) populations of OB stars towards the Ori OB1ab association with distinct distances and radial velocities: Orion~C ($D \sim 412$~pc; $V_{rad} \sim 13$~km~s$^{-1}$) and Orion~D ($D \sim 350$~pc; $V_{rad} \la 5$~km~s$^{-1}$) that lie projected to the north and north-west of Orion~A, respectively \citep[][their Figure~9]{Kounkel2018}. Orion~C has radial velocities similar to the Head of Orion~A, while Orion~D has much lower velocities closer to those of the Tail of Orion~A. An OB star feedback scenario can be imagined where the northern part of the Orion A cloud underwent compressions from shocks driven predominantly by massive stars in Orion~D to produce an extra red-shifted $V_Z$ motion for the entire Head and an extra southward $V_Y$ motion for the northern tip of Orion A.  The compression of the Orion~A by external shocks could have also generated favorable conditions for a global gravitational collapse of the cloud.  Recall that the cluster formation models of \citet{Hartmann2007,Vazquez-Semadeni2017,Vazquez-Semadeni2019} require initial density gradients or colliding streams of gas to initiate global collapse.

{\bf Turbulent clouds} ~~ Molecular line studies have indicated that (magneto)hydrodynamic supersonic turbulence may play an important role in the dynamics of large molecular clouds \citep{MacLow2004,Krumholz2018}.  Turbulence may account, at least in part, for  hierarchical and filamentary structure of cloud complexes, their ``Larson relations'' relating velocity dispersions on different scales, and for their low star formation efficiency.  The curved backbone of the Orion Head is undergoing a twisting motion seen in our 3-D movies (\S\ref{sec_3dmovie}) that may reflect the vorticity expected in turbulent structures.  Furthermore, groups of coeval stars are expected to form by gravitational collapse in localized eddies with small internal relative velocities but larger inter-group relative velocities inherited from the turbulent structure.
	
A giant molecular cloud with long-lived star formation should thus be surrounded by comoving groups of stars with discrepant velocity vectors \citep{Feigelson1996}.  This is seen around the $5-15$~Myr old Sco-Cen Association with several dispersed star groups \citep[e.g., TW Hya Association, $\beta$ Pic Moving Group, $\eta$~Cha cluster;][]{Mamajek1999}.   NGC~1977, and other less populated star groups like the EON and Group~X with discrepant velocities (\S\ref{sec_eon_groupx_discussion}), may represent an earlier stage of the dispersed star groups seen around Sco-Cen.
	
However, some recent studies reevaluate the role of turbulence in clouds suggesting that the observed large widths in molecular gas lines are to be interpreted as self-gravity rather than supersonic turbulent motions \citep[e.g.,][and references therein]{Ballesteros-Paredes2018,Krumholz2018,Vazquez-Semadeni2019}. According to the Global Hierarchical Collapse model of \citet{Vazquez-Semadeni2019}, turbulence is mainly a byproduct of the global gravitational collapse itself; it is relatively weak and its main role is to seed density fluctuations, which in a hierarchical fashion would undergo fragmentations and further gravitational collapses. Echoing some of these ideas, by correcting $^{12}$CO and $^{13}$CO line profiles for the opacity broadening and multiple velocity components effects \citet{Hacar2016a} rule out the presence of supersonic ``small-scale'' turbulence inside the Taurus cloud. 

{\bf Cloud-cloud collision} ~~ Analysis of $^{12}$CO maps lead \citet{Fukui2018} to propose that the formation of the massive OB stars in the Trapezium core and nearby B-type star NU~Ori was triggered by a collision of two clouds: a blue-shifted cloud with $M \sim 15000$~M$_{\odot}$ that includes the bulk of OMC-1 gas (with CO radial velocity $ \sim 8$~km~s$^{-1}$), and a red-shifted cloud with $M \sim 3400$~M$_{\odot}$ (velocity $ \sim 13$~km~s$^{-1}$) that comprises OMC-2/3 and a U-shaped gas structure  surrounding the southern part of OMC-1. 

This U-shaped gas component is noticeable in the gas velocity maps of \citet{Kong2018}, as red-shifted gas patches to south-east and south-west of OMC-1. There is a spatial correlation of our more distant young {\it Gaia} stars in OMC-1 (Figure \ref{fig_distance_vs_declination_head} at $\delta \sim -5.4\degr$; Section \ref{sec_distance}) with the location of these red-shifted gas patches (figure not shown). This correlation qualitatively supports the cloud-cloud collision model providing that the U-shaped red-shifted cloud has already passed the blue-shifted OMC-1 clump.  However, this placement may be inconsistent with the very short collision time-scale of only 0.1~Myr proposed by \citet{Fukui2018}. An alternative explanation for the red-shifted U-shaped component involves expanding CO shells arising from the action of UV radiation and winds from nearby massive stars \citep{Feddersen2018}; specifically, their CO shells \#\# 10 and 11. The cloud-cloud collision model predicts that the massive stars in the Trapezium cluster should be younger than the bulk of the low-mass stars in the ONC; this core-halo age gradient is seen in the ONC \citep[][and Appendix \ref{appendix_b} here]{Getman2014b, Getman2018b}.   

{\bf Slingshot model} ~~ \citet{Stutz2018b} interpret the anti-correlation of stellar distances with gas radial velocities as a signature of a standing wave in the Head of the Orion A cloud, related to their earlier ``slingshot'' model of an oscillating cloud \citep{Stutz2016}. According to this model, the NGC~1977 is being ejected by a whip-like action of the oscillating filament \citep{Stutz2018a}.  However, it is unclear how the same scenario can be applied to clusters in the backbone, such as the ONC, that do not have anomalous motion.  It also does not naturally explain the core-halo age gradient seen in the ONC \citep{Getman2014b}.

\subsection{EON and Group~X high-velocity moving groups}\label{sec_eon_groupx_discussion}  

The EON-north, EON-south, and Group~X stellar groups with discrepant velocities could have formed in any of the scenarios outlined above for NGC~1977.  An additional possibility can be considered for these sparse groups that is unlikely to produce rich clusters like NGC~1977.

{\bf Irradiated cloudlets} ~~ Small clouds on the edges of HII regions can be accelerated by a ``rocket effect'' due to ionization and evaporation of the surfaces facing ionizing OB stars \citep{Oort1955}.  Rocket effect velocities of $1-4$~km~s$^{-1}$ are expected from the simulations of irradiated cloudlets by \citet{Miao2006,Kinnear2015}.

Ionized cloudlets have been detected on the edges of the Extended Orion Nebula cavity.  Using AzTEC mm, NRO $^{12}$CO, and MSX 8~~$\mu$m data, \citet{Shimajiri2011} find molecular cloudlets lying at the edges of the Extended Orion Nebula cavity, which are irradiated by massive stars located in and around the Trapezium cluster. Four elongated structures (designated Regions A, B, C, and D) have their axes directed towards OMC-1, exhibiting spatial stratification characteristic of photodissociation regions. Region A and B cloudlets are located a few arc-minutes to the north and west of our EON-north stellar group, and the Region D cloudlet is positioned $3\arcmin$ to the east of our EON-south stellar group.  Our EONn and EONs stellar groups may have formed in now-dispersed cloudlets $1-2$~Myr ago, inheriting the cloudlet peculiar motions imparted by the rocket effect.

\subsection{S1-East and S3-West Stars}\label{sub_eastweststars}
  
The S1-East and S3-West regions parallel to the main backbone of the Orion~A Head cloud exhibit velocities indicative of slow drifting from the central regions (\S \ref{sec_eastweststars}).  \citet{Kounkel2018} similar found that ONC stars show ``slight preference for expansion near the outer edges''.  In \S \ref{sec_eastweststars}, we suggested this represents portions of a slowly expanding halo of the ONC cluster, or older stars drifting away from star formation along the backbone of the cloud.   This is consistent with literature reports for the presence of distributed stellar populations in numerous star forming regions (e.g., around the Sco-Cen Association, \S\ref{sec_ngc1977_head_discussion}). In MYStIX and SFiNCs studies of rich clusters in massive molecular clouds, dispersed young stellar populations surrounding the compact clusters are ubiquitous on spatial scales of $5-20$~pc \citep{Kuhn2014, Getman2018b}. These distributed populations have older ages than the main MYStIX and SFiNCs clusters \citep{Getman2014a,Getman2018b}, suggesting continuous or episodic star formation in massive molecular clouds for many millions of years.  A similar process of slow dispersal of recently formed stars around the Orion~A cloud is indicated by the S1-East and S3-West populations.  Numerous members of an older dispersed population around the Orion complex have been found by \citet{Carpenter2001}. 


\section{Conclusions}\label{sec_conclusions}

We collect a rich sample of $\sim 1500$ previously published optical, infrared, and X-ray selected stars across the Orion~A region that have reliable pre-main sequence ages, ground-based measurements of radial velocities, and proper motions and parallaxes measured with the {\it Gaia} satellite (\S\S \ref{sec_starsamples} and \ref{sec_radial_velocities}). Our stellar kinematic analyses are performed with respect to the rest frame of the star center in the Head (OMC-1/2/3/4/5) of Orion A using a Cartesian $x$, $y$, $z$ coordinate system where the $x$ and $y$ axes are orthographic projections of the $-\alpha$ and $+\delta$ celestial lines, and the $z$-axis is directed along the line of sight (\S\ref{sec_stellarkinematics}). Based on the previously published findings of strong correlation between star and gas radial velocities and its changes across individual main parts of the cloud's Head, main  kinematic stellar structures are chosen here to be associated with the main molecular components, such as OMC-2/3, OMC-1, OMC-4, and OMC-5 (\S \ref{sec_stellargroups}). 

Our analyses of the adaptive smoothed kinematics/distance maps and position-velocity diagrams for the cloud's Head led to an identification of additional star cluster and groups with peculiar kinematics (\S \ref{sec_stellargroups}). These include sparse  groups of stars to the east (S1-East) and west (S3-West) of the cloud, the relatively rich cluster NGC~1977,  a star group south of NGC~1977 (NGC~1977-south), two groups in the Extended Orion Nebula cavity (EON-n and EON-s), and a Group-X to the west of the cloud (\S\S \ref{sec_ngc1977} and \ref{sec_eastweststars}).  These stellar structures are not embedded in the cloud and have estimated ages between 1.5 and 3~Myr.  

We confirm previously reported findings of likely varying distances from the observer to different parts of the cloud's Head (\S \ref{sec_distance}) and the negative correlations between gas velocities and Gaia distances (\S \ref{sec_relationship_d_vs_vr}).  With a larger sample and distinct kinematic groups, we produce movies of the 3-dimensional motions of various star structures (\S \ref{sec_3dmovie}).  The movies demonstrate contraction of stars within OMC-1 and a more  global contraction of OMC-1 and OMC-2/3 stars.  Gravitational infall is thus confirmed to be present along the Orion Head backbone.  

In \S \ref{sec_discussion}, we discuss various scenarios for the origin of the cluster and groups with peculiar kinematics.  The unusual motion of the NGC~1977 cluster may be linked to large-scale mechanisms including global hierarchical gravitational cloud collapse, irradiation feedback from massive stars, turbulence, cloud-cloud collision, or a ``slingshot'' process.  The sparse EON and Group~X groups may also have acquired anomalous velocities due to small-scale ``rocket effects''.  The older S1-East and S3-West groups can be viewed as portions of a slowly expanding ONC halo or stars slowly drifting from their origins in the Orion~A central filament. 

While the ``global hierarchical gravitational collapse'' and ``feedback by OB stars'' are favored by us, the other scenarios are not ruled out. But some broad understanding emerges.  The findings give the strong impression that, while gravitational infall is present, no single, simple process dominates the kinematics of young stars on these multi-parsec scales.  More complex processes such as OB star feedback, supersonic turbulence,  cloud collisions or magnetically driven instabilities may be active in Orion~A, the nearest giant molecular cloud.  Both the gravitational contraction of OMC 1/2/3 stars and the anomalous kinematics of outlying star groups probably reflect motions of their natal cloud substructures.  Gravitational gas contraction is seen along the contemporary gas filament, possibly with turbulent eddies superposed; similar motions were likely present in gas now dissipated that gave rise to older star groups.  A unified view of young stellar structures and kinematics, gas and dust structure and kinematics, and the astrophysics of collapsing large molecular clouds is emerging.



\section*{Acknowledgements}

We thank the referee J. Alves for many useful suggestions that helped to improve this work. We thank J. Forbrich and T. Prusti for stimulating discussions. The MYStIX project is now supported by the {\it Chandra} archive grant AR7-18002X. The SFiNCs project is supported at Penn State by NASA grant NNX15AF42G, and the {\it Chandra} ACIS Team contract SV474018 (G. Garmire \& L. Townsley, Principal Investigators), issued by the {\it Chandra} X-ray Center, which is operated by the Smithsonian Astrophysical Observatory for and on behalf of NASA under contract NAS8-03060. The Guaranteed Time Observations (GTO) data used here were selected by the ACIS Instrument Principal Investigator, Gordon P. Garmire, of the Huntingdon Institute for X-ray Astronomy, LLC, which is under contract to the Smithsonian Astrophysical Observatory; Contract SV2-82024. This research has made use of NASA's Astrophysics Data System Bibliographic Services and SAOImage DS9 software developed by Smithsonian Astrophysical Observatory.



\bibliographystyle{mnras}
\interlinepenalty=1000
\bibliography{Bibliography} 

\appendix
\section{Three catalog tables}\label{appendix_3tables}

This section provides information on formatting and content of three tables that list catalogs of young stars in Orion A employed in the current paper. Table \ref{tbl_gaia_props} focuses on {\it Gaia} properties of 1487 young stars, including position, parallax, proper motions, and a flag indicating photometric BP/RP flux excess that allows division into two samples, \texttt{full} and \texttt{restricted} (see details in \S~\ref{sec_starsamples}). 

For these 1487 young stars, Table \ref{tbl_other_props} lists additional properties taken from previous literature. Column 2 indicates if the {\it Gaia} source is part of one of the aforementioned six optical/IR/X-ray catalogues of young stars. Among the 1487 {\it Gaia} stars, 528, 532, 633, 357, 1185, and 568 are in MYStIX/SFiNCs \citep{Broos2013,Getman2017}, \citet{Furesz2008}, \citet{Tobin2009},  \citet{DaRio2012}, \citet{DaRio2016}, and \citet{Megeath2012}, respectively. Columns 3 and 4 give $J$ and $H$-band magnitudes from \citet{Meingast2016}. Column 5 lists visual sources extinction estimate ($A_V$), either directly taken from the catalog of IR bright stars from  \citet{DaRio2016} when available or derived from the $A_V$ - $(J-H)$ relationship assuming the source is a low-mass star. Columns 6 and 7 provide two types of age estimates, the traditional HRD estimate based on optical/IR photometry/spectroscopy from \citet{DaRio2016} and our $Age_{JX}$ estimate based on X-ray and near-IR photometry \citep{Getman2014a,Getman2017}; the latter is only applicable to low-mass stars. Both types of ages are corrected for individual {\it Gaia} parallaxes and calibrated to isochrones by  \citet{Siess2000}. There is a good agreement between these two age estimates (\S \ref{appendix_b}). Column 8 gives apparent mid-IR SED (spectral energy distribution) slope inferred from the {\it Spitzer}-IRAC point source catalog of \citet[][publicly available online]{Megeath2012}. The last column lists a flag indicating membership in various stellar structures of interest discussed in this paper.

For 2752 young stars, Table~\ref{tbl_vrad} lists heliocentric average stellar velocity measurements from Kounkel et al. and Tobin et al. along with $J$ and $H$-band magnitudes from \citet{Meingast2016}. Velocities from Tobin et al. (Column 5) are listed only for stars that are not present in the Kounkel et al. APOGEE catalog.  Column 3 indicates that among the 2752 stars, 806, 948, 1200, 600, 2275, 1003, and 1192 are in MYStIX/SFiNCs \citep{Broos2013,Getman2017}, \citet{Furesz2008}, \citet{Tobin2009},  \citet{DaRio2012}, \citet{DaRio2016}, \citet{Megeath2012}, and our {\it Gaia}-selected young star catalogs (Table~\ref{tbl_gaia_props}), respectively.

\begin{table*}\normalsize
	\centering
	\begin{minipage}{180mm}
		\caption{{\it Gaia}-selected catalog of young stars in Orion A; {\it Gaia} properties.  This table is available in its entirety (1487 stars) in the machine-readable form in the online journal. A portion is shown here for guidance regarding its form and content. Column 1: {\it Gaia} DR2 source ID. Columns 2-3: Right ascension and declination for epoch J2000.0 in degrees. Columns 4-6: {\it Gaia} parallax and proper motion. Only stars with statistical error on  $\sigma_{\bar{\omega}}<0.1$~mas are included in this catalog. Column 7: $G$-band mean magnitude. Column 8: Flag indicating {\it Gaia} photometric BP$/$RP flux excess. The flag is set to ``1'' when the flux excess is small, $(I_{BP}+I_{RP})/I_{G} < 1.35 + 0.06 \times (G_{BP}-G_{RP})^{2}$ \citep{Evans2018}; otherwise the flag is set to ``0''.}
		\label{tbl_gaia_props}
		\begin{tabular}{@{ \vline }c@{ \vline }c@{ \vline }c@{ \vline }c@{ \vline }c@{ \vline }c@{ \vline }c@{ \vline }c@{ \vline }}
			\cline{1-8}
			&&&&&&&\\
			ID & R.A. & Decl. & ${\bar{\omega}}$ & $\mu_{\alpha\star}$ & $\mu_{\delta}$ & $G$ & FEC\\ 
			&(deg)&(deg)&(mas)&(mas~yr$^{-1}$)&(mas~yr$^{-1}$)& (mag) &\\
			(1)&(2)&(3)&(4)&(5)&(6)&(7)&(8)\\
			\cline{1-8}
			&&&&&&&\\
			3017252291090540288 &    83.894572 &    -5.577275 & $ 2.475\pm 0.095$ & $   1.139\pm   0.143$ & $   0.556\pm   0.116$ & $  15.639\pm   0.003$ &       0\\
			3017252428529514752 &    83.796422 &    -5.614225 & $ 2.636\pm 0.069$ & $   1.464\pm   0.125$ & $  -0.223\pm   0.098$ & $  14.120\pm   0.004$ &       1\\
			3017252462889247744 &    83.831229 &    -5.614398 & $ 2.632\pm 0.090$ & $   1.106\pm   0.157$ & $   0.448\pm   0.130$ & $  15.389\pm   0.002$ &       1\\
			3017252600328207104 &    83.781716 &    -5.605461 & $ 2.335\pm 0.054$ & $   0.950\pm   0.095$ & $  -3.892\pm   0.078$ & $  13.740\pm   0.006$ &       1\\
			3017252600328208384 &    83.784268 &    -5.617997 & $ 2.465\pm 0.063$ & $   1.695\pm   0.109$ & $  -1.353\pm   0.099$ & $  15.273\pm   0.011$ &       0\\
			3017252664749740160 &    83.771132 &    -5.612182 & $ 2.415\pm 0.087$ & $   1.625\pm   0.163$ & $  -0.094\pm   0.137$ & $  16.200\pm   0.015$ &       0\\
			3017252772126891392 &    83.814545 &    -5.586646 & $ 2.354\pm 0.094$ & $   0.994\pm   0.169$ & $  -0.128\pm   0.140$ & $  16.090\pm   0.003$ &       0\\
			3017252806486631552 &    83.796367 &    -5.583240 & $ 2.201\pm 0.097$ & $   1.103\pm   0.169$ & $  -0.623\pm   0.136$ & $  15.482\pm   0.003$ &       0\\
			3017252943925578752 &    83.840406 &    -5.579685 & $ 2.433\pm 0.032$ & $   0.966\pm   0.056$ & $  -0.543\pm   0.048$ & $  13.234\pm   0.015$ &       1\\
			3017252943925579008 &    83.840098 &    -5.582887 & $ 2.335\pm 0.059$ & $   2.292\pm   0.108$ & $   0.122\pm   0.101$ & $  15.019\pm   0.007$ &       1\\
			&&&&&&&\\
			\cline{1-8} 
		\end{tabular}
	\end{minipage}
\end{table*}

\begin{table*}\normalsize
	\centering
	\begin{minipage}{180mm}
		\caption{{\it Gaia}-selected catalog of young stars in Orion A; Other Properties. This table is available in its entirety (1487 stars) in the machine-readable form in the online journal. A portion is shown here for guidance regarding its form and content. Column 1: {\it Gaia} DR2 source ID. Column 2: Six-digit flag indicating the presence of the {\it Gaia} source in the following X-ray, optical, infrared source catalogs. First digit $=1$ indicates the source is present in the MYStIX \citep{Broos2013} and/or SFiNCs \citep{Getman2017} catalogs; second, third, forth, fifth, and sixth digits $=1$ indicate the source is in the following optical/infrared catalogs, respectively: \citet{Furesz2008, Tobin2009, DaRio2012, DaRio2016, Megeath2012}. The sixth digit here is applicable only to the \citet{Megeath2012} catalog of young stellar objects in Orion A, whereas Column 8 below is related to the entire {\it Spitzer} catalog of point sources by Megeath et al. Columns 3-4: $J$ and $H$-band magnitudes from \citet{Meingast2016}. Column 5: Visual source extinction based on the  extinction scale from \citet{DaRio2016}. Columns 6: Stellar age estimate from \citet{DaRio2016} corrected here for individual stellar parallax and placed on the \citet{Siess2000} time-scale. Column 7: $Age_{JX}$ from MYStIX \citep{Getman2014a} and SFiNCs \citep{Getman2017} corrected for individual stellar parallax and placed on the \citet{Siess2000} time-scale. Column 8: Apparent MIR SED (spectral energy distribution) slope measured in the {\it Spitzer}-IRAC wavelength range from 3.6 to 8.0 $\mu$m as $\alpha_{IRAC} = d \log(\lambda F_{\lambda})/d \log(\lambda)$, using IRAC data from the extended {\it Spitzer} catalog of point sources in Orion A/B by \citet{Megeath2012}. Column 9: Membership flag: ``1'', ``2'', ``3'', ``4'', and ``5'' - stars in the main NGC~1977 cluster, NGC~1977 south group, EON north and EON south, and Group~X groups, respectively; and ``0'' -  stars outside those cluster/groups.}
		\label{tbl_other_props}
		\begin{tabular}{@{ \vline }c@{ \vline }c@{ \vline }c@{ \vline }c@{ \vline }c@{ \vline }c@{ \vline }c@{ \vline }c@{ \vline }c@{ \vline }c@{ \vline }}
			\cline{1-9}
			&&&&&&&&\\
			ID & Cat & $J$ & $H$ & $A_V$ & $Age_{D16}$ & $Age_{JX}$ & $\alpha_{IRAC}$ & Mem\\ 
			&&(mag)&(mag)&(mag)&(Myr)&(Myr)&&\\
			(1)&(2)&(3)&(4)&(5)&(6)&(7)&(8)&(9)\\
			\cline{1-9}
			&&&&&&&&\\ 
			3017252291090540288 & 111110 & $12.831\pm 0.026$ & $12.117\pm 0.003$ &   0.6 &   2.4 & 2.5 & ... & 0\\
			3017252428529514752 & 100110 & $11.124\pm 0.024$ & $10.267\pm 0.032$ &   1.2 &   0.8 & 1.1 & $-2.93\pm 0.01$ & 0\\
			3017252462889247744 & 100110 & $12.499\pm 0.029$ & $11.822\pm 0.003$ &   0.4 &   2.1 & 2.5 & $-2.72\pm 0.02$ & 0\\
			3017252600328207104 & 100110 & $10.896\pm 0.024$ & $10.080\pm 0.034$ &   1.3 &   0.1 & 0.6 & $-2.57\pm 0.01$ & 0\\
			3017252600328208384 & 111111 & $11.794\pm 0.024$ & $10.872\pm 0.030$ &   1.6 &   1.0 & 1.9 & $-1.03\pm 0.01$ & 0\\
			3017252664749740160 & 111111 & $12.729\pm 0.024$ & $11.909\pm 0.032$ &   1.5 &   1.1 & 3.0 & $-1.17\pm 0.01$ & 0\\
			3017252772126891392 & 111110 & $12.705\pm 0.023$ & $11.790\pm 0.033$ &   2.0 &   1.4 & 1.7 & $-2.40\pm 0.03$ & 0\\
			3017252806486631552 & 100110 & $11.948\pm 0.023$ & $11.089\pm 0.030$ &   1.8 &   0.3 & 1.6 & $-2.54\pm 0.01$ & 0\\
			3017252943925578752 & 111111 & $11.181\pm 0.023$ & $10.319\pm 0.032$ &   1.1 &   0.7 & 0.7 & $-0.45\pm 0.00$ & 0\\
			3017252943925579008 & 111111 & $12.252\pm 0.024$ & $11.418\pm 0.032$ &   1.2 &   1.3 & 1.2 & $-1.33\pm 0.01$ & 0\\
			&&&&&&&&\\
			\cline{1-9}  
		\end{tabular}
	\end{minipage}
\end{table*}

\begin{table*}\normalsize
	\centering
	\begin{minipage}{180mm}
		\caption{Catalog of young stars with radial velocities in Orion A. Velocity data are taken from Table~1 of \citet{Kounkel2018} and Table~3 of \citet{Tobin2009}. This table is available in its entirety (2752 stars) in the machine-readable form in the online journal. A portion is shown here for guidance regarding its form and content. Columns 1-2: Right ascension and declination for epoch J2000.0 in degrees. Column 3: Seven-digit flag indicating the presence of the source in the following X-ray, optical, infrared catalogs of young stars. First digit $=1$ indicates the source is present in the MYStIX \citep{Broos2013} and/or SFiNCs \citep{Getman2017} catalogs; second, third, forth, fifth, and sixth digits $=1$ indicate the source is in the following optical/infrared catalogs, respectively: \citet{Furesz2008, Tobin2009, DaRio2012, DaRio2016, Megeath2012}. Seventh digit $=1$ indicates that the source is included in the {\it Gaia} catalog listed in Tables~\ref{tbl_gaia_props} and \ref{tbl_other_props}. The sixth digit here is applicable only to the \citet{Megeath2012} catalog of young stellar objects in Orion A.  Column 4: Average heliocentric radial velocity from \citet{Kounkel2018}. Columns 5-6: Average heliocentric radial velocity and spectroscopic binary flag from \citet{Tobin2009}; given here only for young stars without Kounkel et al. velocity measurements. Columns 7-8: $J$ and $H$-band magnitudes from \citet{Meingast2016}. Column 9: Membership flag: ``1'', ``2'', ``3'', ``4'', and ``5'' - stars in the main NGC~1977 cluster, NGC~1977 south group, EON north and EON south, and Group~X groups, respectively; and ``0'' -  stars outside those cluster/groups.}
		\label{tbl_vrad}
		\begin{tabular}{@{ \vline }c@{ \vline }c@{ \vline }c@{ \vline }c@{ \vline }c@{ \vline }c@{ \vline }c@{ \vline }c@{ \vline }c@{ \vline }c@{ \vline }}
			\cline{1-9}
			&&&&&&&&\\
			R.A. & Decl. & Cat & $\overline{V_{rad;K18}}$ & $\overline{V_{rad;T09}}$ & Mult & $J$ & $H$ & Mem\\ 
			(deg)&(deg)&&(km~s$^{-1}$)&(km~s$^{-1}$)&&(mag)&(mag)&\\
			(1)&(2)&(3)&(4)&(5)&(6)&(7)&(8)&(9)\\
			\cline{1-9}
			&&&&&&&&\\ 
			83.787476 &    -4.170634 & 0000100 & $  30.889\pm   0.125$ & ... & ... & $10.746\pm 0.022$ & $10.186\pm 0.030$ & 0\\
			83.904503 &    -4.168316 & 0000100 & $   3.822\pm   0.334$ & ... & ... & $11.811\pm 0.023$ & $11.425\pm 0.030$ & 0\\
			84.185982 &    -4.153669 & 0000100 & $  21.291\pm   0.111$ & ... & ... & $11.080\pm 0.023$ & $10.360\pm 0.031$ & 0\\
			83.970421 &    -4.135372 & 0000111 & $  26.685\pm   0.383$ & ... & ... & $13.124\pm 0.004$ & $12.471\pm 0.003$ & 0\\
			83.956345 &    -4.125016 & 0000100 & $ -41.196\pm   0.492$ & ... & ... & $13.181\pm 0.005$ & $12.471\pm 0.003$ & 0\\
			83.344087 &    -5.544464 & 0110001 & ... & $   23.9\pm    1.0$ &   1 & $13.033\pm 0.003$ & $12.303\pm 0.002$ & 0\\
			83.347171 &    -5.364753 & 0110000 & ... & $   47.2\pm    0.1$ &   1 & $12.904\pm 0.023$ & $12.296\pm 0.003$ & 0\\
			83.356725 &    -5.398375 & 0110001 & ... & $   69.9\pm    0.8$ &   1 & $13.319\pm 0.004$ & $12.595\pm 0.003$ & 0\\
			83.368196 &    -4.930408 & 0110000 & ... & $  -22.1\pm    0.9$ &   1 & $13.320\pm 0.003$ & $12.606\pm 0.002$ & 0\\
			83.369842 &    -5.436131 & 0110001 & ... & $   29.2\pm    0.8$ &   1 & $13.185\pm 0.004$ & $12.437\pm 0.003$ & 0\\
			&&&&&&&&\\
			\cline{1-9}  
		\end{tabular}
	\end{minipage}
\end{table*}

\section{Transformation of proper motions and radial velocities to $V_X$, $V_Y$, and $V_Z$}\label{appendix_v_transformation}

Recall from \S\ref{sec_starsamples} that our analyses are carried out for two {\it Gaia} star subsamples, \texttt{full} and \texttt{restricted}. The latter, a more conservative approach, trims 266 stars with high BP/RP flux excess; the vast majority of these stars reside within the central part of ONC around the OMC-1 filament where the effects of background nebular emission and source crowding are the highest. Average kinematic properties of the star center differ slightly for the two {\it Gaia} subsamples by about 0.01 mas in parallax and 0.4 mas yr$^{-1}$ in proper motion.  For the \texttt{full} permutation, the mean motion and distance values are $\mu_{\alpha{\star,0}} = 1.459$~mas~yr$^{-1}$, $\mu_{\delta,0}=0.646$~mas~yr$^{-1}$, $V_{rad,0} = 26.6$~km~s$^{-1}$, and ${\bar{\omega_0}}=2.529$~mas. For the \texttt{restricted} permutation, the mean motion and distance values are $\mu_{\alpha{\star,0}} = 1.645$~mas~yr$^{-1}$, $\mu_{\delta,0}=0.306$~mas~yr$^{-1}$, $V_{rad,0} = 26.6$~km~s$^{-1}$, and ${\bar{\omega_0}}=2.543$~mas.

In order to estimate stellar velocities $V_X$, $V_Y$, and $V_Z$, following \citet{Kuhn2018} we perform four main transformations. First, we calculate proper motion terms to correct for the effect of perspective contraction or expansion; that is, star clusters appear to contract or expand as they move away from or towards the observer \citep{vanLeeuwen2009,GCHelmi2018}. This is achieved by applying formulas (1) and (2) from Kuhn et al. or formula (13) from \citet{vanLeeuwen2009} that involve eight independent variables. Second, following formulas (3) and (4) from Kuhn et al., the apparent star proper motions are corrected for the perspective contraction/expansion and the motion of the star center, and are converted to velocities in the $-\alpha$ and $+\delta$ plane. Third, the latter velocities are transformed to the $x$ and $y$ plane following the orthographic projection formulas (2) from \citet{GCHelmi2018}, which involve 6 independent variables. Fourth, the observed stellar radial velocities are transformed to $V_Z$ velocities by subtracting $V_{rad,0}$.

\section{Raw spatial maps of distance and velocities across the Head of Orion A}\label{appendix_raw_maps}

\begin{figure*}
	\includegraphics[angle=0.,width=160mm]{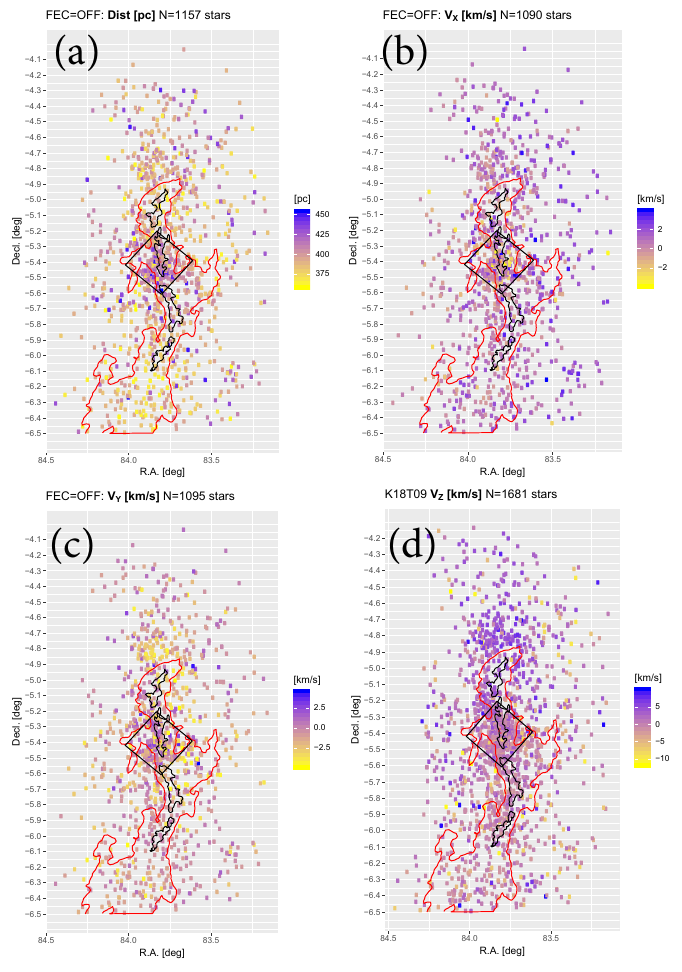}
	\caption{Raw spatial maps of stellar distances $D$  and velocity components, $V_X$, $V_Y$, and $V_Z$ for the {\it Gaia}  \texttt{full} sample.   The $V_Z$ measurements are given for the ``K18T09'' sample. Outliers with extreme values of $D$, $V_X$, $V_Y$, and $V_Z$ are excluded from these maps to shorten dynamic ranges and allow meaningful color scales. For the OMC-12345 gas components, the C$^{18}$O$(1-0)$ emission contours at 7~K~km~s$^{-1}$ from \citet{Kong2018} are in black. As reference contours, the MYStIX-COUP field of view (black square) and the $A_V = 6$~mag contour (red) are also provided. Figure panels showing the $D$, $V_X$, $V_Y$ maps for the {\it Gaia} \texttt{restricted} sample are provided in the Supplementary Materials of Figure \ref{fig_smoothed_maps}. The Supplementary Materials also include a $V_Z$ panel for the ``{\it Gaia}-K18T09'' star sample.}
	\label{fig_raw_maps}
\end{figure*}

Given here is Figure~\ref{fig_raw_maps} showing raw (including individual stars) spatial maps of {\it Gaia} distance and three velocity components across the Head of Orion A. Each map comprises over 1000 young stellar members of the region. This figure complements Figure \ref{fig_smoothed_maps}, which provides adaptively smoothed maps of these quantities.

\section{Visualizing Core-Halo Age Gradient In ONC}\label{appendix_b}

Using the $Age_{JX}$ estimator of PMS stellar ages derived from X-ray and near-IR photometry of $M \la 1.2$~M$_{\odot}$ pre-main sequence stars \citet{Getman2014a, Getman2014b} report core-halo age gradients with younger cores and older halos in the ONC and NGC~2024 clusters. These findings are independently supported by the spatial gradients in the disk fraction and $K_s$-band excess frequency. \citet{Getman2018a} further demonstrate that such core-halo age gradients are generally present in young, rich, isolated MYStIX$+$SFiNCs clusters. These findings of late or continuing star formation in the cores of clusters with older stars dispersed in the outer regions are in line with the predictions of the global hierarchical collapse model by \citet{Vazquez-Semadeni2017,Vazquez-Semadeni2019}.

\begin{figure*}
	\includegraphics[angle=0.,width=180mm]{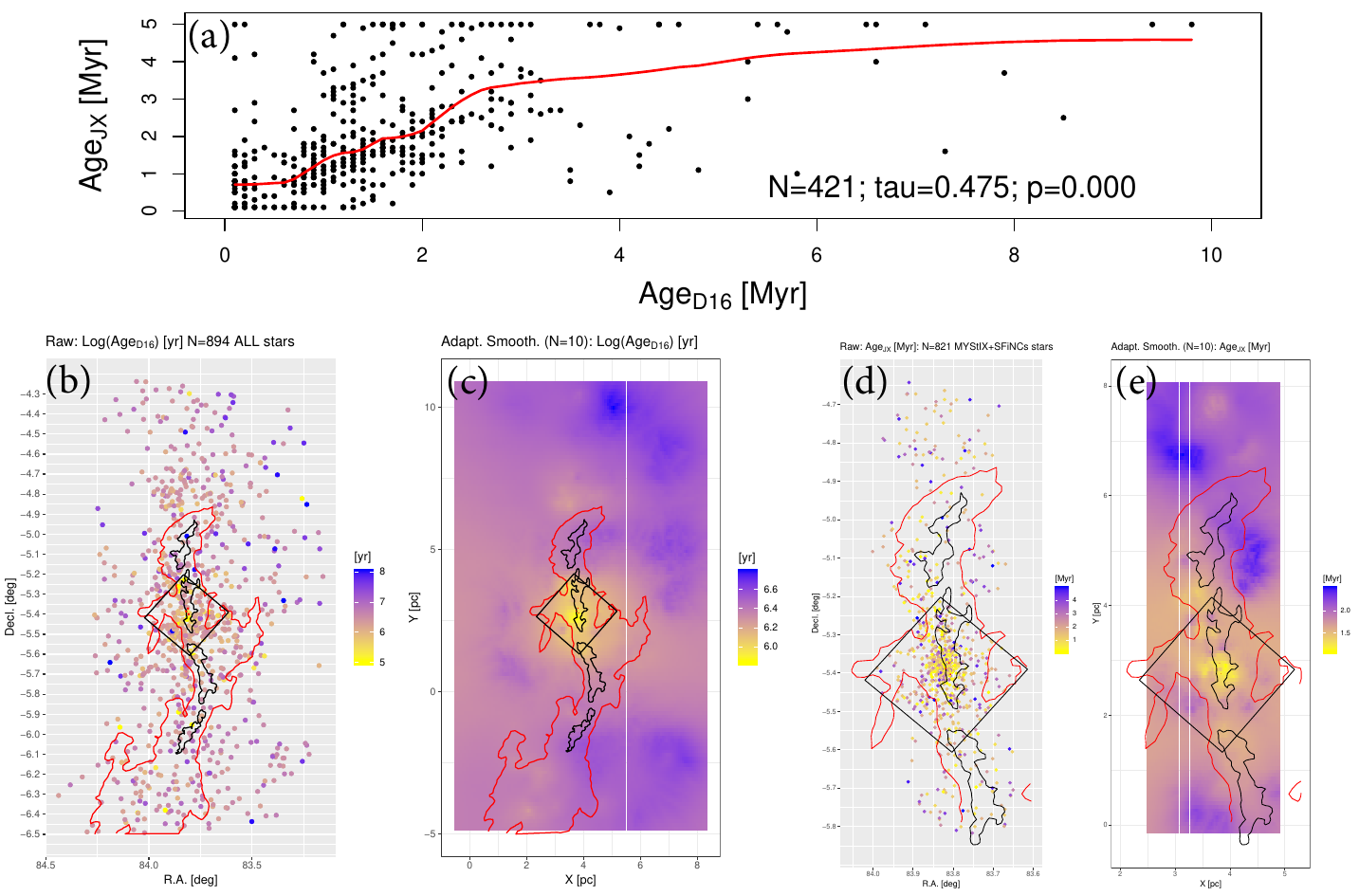}
	\caption{Ages for {\it Gaia}-selected stars in the Head of Orion A.  Panel a: Comparison of $Age_{JX}$ \citep{Getman2014a} with ages from \citet{DaRio2016}, both corrected for individual {\it Gaia} stellar parallaxes. The legend states number of stars, Kendall's $\tau$ correlation coefficient and its corresponding $p$-value. A local quadratic regression fit is shown in red. Panels b and c:  Individual star and adaptively smoothed maps of age from \citet{DaRio2016}. Panels d and e: Individual star and adaptively smoothed maps of $Age_{JX}$ for the entire (regardless of {\it Gaia} selection) sample of MYStIX+SFiNCs X-ray stars with available $Age_{JX}$ estimates. On all panels, the three tiny red points mark the locations of $\Theta^{1}$Ori~C, BN-KL and OMC-1S subregions.}
	\label{fig_age_maps}
\end{figure*}

Throughout the current paper, we opt to employ traditional age estimates, derived from optical-infrared spectroscopy/photometry data, provided by \citet{DaRio2016}. The main reason for omitting $Age_{JX}$ is that such estimates are not available for our {\it Gaia} stars located outside the MYStIX$+$SFiNCs fields and/or for stars with $M \ga 1.2$~M$_{\odot}$.  Figure \ref{fig_age_maps}a shows that for {\it Gaia} stars that are common between \citet{DaRio2016} and MYStIX$+$SFiNCs the two age estimates are consistent with each other, though with considerable scatter. 

Figures~\ref{fig_age_maps}(d,e) show raw and adaptively smoothed maps of $Age_{JX}$ for the entire sample of MYStIX$+$SFiNCs stars to visualize the core-halo age gradient in ONC. Notice that in their age analysis of ONC \citet{Getman2014b}  use only lightly absorbed stars ($A_V \la 5$~mag) to reduce possible contamination from stellar populations embedded in the cloud; but since $Age_{JX}$ is not sensitive to protostars and the smoothed maps for the lightly and heavily absorbed stars look similar (not shown), here we retain the entire MYStIX$+$SFiNCs star sample with available $Age_{JX}$ estimates. The purpose of these maps is to emphasize that the younger stars lie in and near the Trapezium cluster.

For our {\it Gaia} star sample (Table \ref{tbl_other_props}), Figures~\ref{fig_age_maps}(b,c) show raw and adaptively smoothed maps of age estimates from \citet{DaRio2016}. This {\it Gaia} sample generally excludes members of the Trapezium cluster owing to higher obscuration and the presence of high nebular emission in this region.  This data selection effect results in an apparent core-halo age gradient with the minimum ages shifted to the south of Trapezium.
 
\bsp	
\label{lastpage}
\end{document}